\documentclass [11pt]{article}
\usepackage{graphicx}
\usepackage{psfrag}
\usepackage{amsmath,amssymb}
\makeatletter
\def\section{\@startsection {section}{1}{\z@}{-3.5ex plus -1ex minus
 -.2ex}{2.3ex plus .2ex}{\large\bf}}
\def\subsection{\@startsection{subsection}{2}{\z@}{-3.25ex plus -1ex
minus -.2ex}{1.5ex plus .2ex}{\normalsize\bf}}
\makeatother
\makeatletter

\@addtoreset{equation}{section}

\makeatother

\def\dslash{\raisebox{1pt}{$\slash$} \hspace{-7pt} \partial}

\def\bea{\begin{eqnarray}} \def\eea{\end{eqnarray}}
\def\be{\begin{equation}} \def\ee{\end{equation}} \def\nn{\nonumber}
\def\a{& \hspace{-7pt}}  \def\Z{{\bf Z}}
 \def\ov{\overline} \def\I{1\hspace{-4pt}1}
\def\ds{\displaystyle}  
 
  \def\RR{\mathbb R}
\def\mc{\mathcal}
\def\ZZ{Z\hspace{-5pt}Z}
\def\CC{\mathbb C}

\newcommand\tw[2]{
 \Bigg[\hspace{-1pt}\raisebox{1pt}
 {$\begin{array}{c}
 \displaystyle{#1} \\ \displaystyle{#2}
 \end{array}$}
 \hspace{-1pt}\Bigg]}

\begin{document}

\thispagestyle{empty}

\begin{center}
\hfill SISSA-64/2004/EP \\

\begin{center}

\vspace{1.7cm}

{\LARGE\bf Orbifold resolutions and \\[3mm]
fermion localization}

\end{center}

\vspace{1.4cm}

{\bf Marco Serone and  Andrea Wulzer}\\

\vspace{1.2cm}

{\em ISAS-SISSA and INFN, Via Beirut 2-4, I-34013 Trieste, Italy}
\vspace{.3cm}

\end{center}

\vspace{0.8cm}

\centerline{\bf Abstract}
\vspace{2 mm}
\begin{quote}\small

We study the Dirac equation of chiral fermions on a regularized version
of the two-dimensional $T^2/\Z_2$ orbifold, where the conical
singularities are replaced by suitable spherical caps with constant
curvature. This study shows how localized and bulk fermions arise in
the orbifold as the resolved space approaches the orbifold limit.

Our analysis also shows that not all possible fermion configurations on $T^2/\Z_2$
admit such a simple resolution. We focus our study to a fermion coupled to a $U(1)$ gauge field.
It is explicitly shown how a resolution of the orbifold puts severe constraints
on the allowed chiralities and $U(1)$ charges of the massless four dimensional
fermions, localized or not, that can be present in the orbifold.

The limit in which $T^2/\Z_2$ (and its corresponding resolved space)
collapses to $S^1/\Z_2$ is also studied in detail.
\end{quote}

\vfill

\newpage

\section{Introduction}

In the last years there has been an intense study of theories in extra dimensions
from a bottom-up approach. These works are all motivated by the search of
a framework or model that explains some of the open theoretical issues of the Standard Model (SM)
(such as the hierarchy problem \cite{RS12}),
and/or it provides a more elegant and unified framework (such as gauge unification \cite{gauge-uni})
and/or it has less free parameters, providing some predictions (such as the value
of the Higgs mass \cite{BHN}). From a phenomenological point of view, this interest
is further motivated by the search of an alternative to the standard scenario of physics beyond
the SM, namely low energy Supersymmetry (SUSY).
The lack of the discovery of super partners starts to pose a fine-tuning problem
in the Minimal Supersymmetric Standard Model (MSSM), motivating the investigation of other possible
scenarios for physics beyond the SM, including extra dimensions at the
${\rm TeV}$ scale \cite{tev}. From a more theoretical point of view, extra dimensions
are strongly motivated (and somehow ``predicted'') by string theory, which in flat space requires
the existence of 10 (or 11) space-time dimensions.
It is, however, only after the appearance of ref.~\cite{HW} that field theories in extra
dimensions became fashionable also outside the string community and started to be taken
seriously by a wider range of theoretical physicists.

In the context of higher dimensions, a particular useful construction is provided by
orbifold compactifications \cite{Dixon}. Orbifolds are singular and yet tractable spaces
which provide an easy way (maybe the easiest) to generate
four-dimensional chirality and break higher-dimensional symmetries.
As a matter of fact, indeed, most of the recent work on (compact) higher-dimensions
is based on simple orbifold models, typically $S^1/\Z_2$ or $T^2/\Z_2$,
respectively for five dimensional (5D) and six dimensional (6D) theories.
The current rules in constructing orbifold field theories are quite flexible.
One can introduce an arbitrary number of fields propagating along the
extra dimensions (bulk fields) or localized at the
orbifold fixed points (localized fields). The property and number of both
bulk and localized fields is arbitrary, provided that they respect the underlying global and
local symmetry of the system, and lead to an anomaly free model.
This flexibility and simplicity is one of the main reasons explaining why the field-theoretical approach
to orbifolds is currently fashionable.
One should not forget, however, that orbifolds are singular spaces
and, as such, they should always be seen as limits of smooth spaces.
String-theory on orbifolds, for instance, is well-behaved and free of any
kind of inconsistency or singularity because it is understood that such models
arise as particular limits of string compactifications on smooth spaces.
One can also spot the fields whose vacuum expectation values provide
the resolution of the orbifold singularity.

Aim of this work is to study how and whether field theory orbifolds can be seen as
particular limits of some smooth spaces and how bulk and localized fields
arise in this limit. For simplicity, we will consider only the dynamics of fermions coupled
to a $U(1)$ gauge field, on conical singularities of the form $\CC/\Z_N$. We
study in detail the compact orbifold $T^2/\Z_2$ and its degeneration limit leading to
an $S^1/\Z_2$ orbifold.

Massless four dimensional chiral fermions are in one-to-one correspondence with the number and chirality
of the massless two-dimensional chiral fermions on the compact space ${\cal M}$. We are then led
to study the massless Dirac equation on ${\cal M}$.
This analysis would require to find a family of two-dimensional smooth compact manifolds
({\it i.e}. Riemann surfaces) which, in a suitable limit, gives rise to the orbifold $T^2/\Z_2$,
and in which the Dirac equation is tractable. We have not been able to find
a smooth and tractable family of surfaces of this sort. Instead, we have used a ``cut and paste''
regularization of the conical singularities consisting in cutting a small portion of the
cone that includes its apex, and in replacing it with a suitable portion of spherical cap, such that
the resulting space is endowed with a continuous tangent space. In this way, we replace
the $T^2/\Z_2$ orbifold, that locally has four conical singularities of the kind $\CC/\Z_2$,
with four spherical caps and a flat region connecting them.
The resulting space ${\cal M}$ is not totally smooth, giving rise to a discontinuous curvature, yet
it provides a resolution of the orbifold singularity.\footnote{To be precise, the words
``resolution'' or ``deformation'' are actually improper, since i) they are typically referred to
complex manifolds where a complex structure is preserved and ii) because 2D conical singularities
are actually not orbifold singularities. A better name would be ``unfolding'' the singularity.
Higher dimensional spaces, such as $\CC^n/Z_N$ and their compact versions $T^n/\Z_N$ ($n>1$),
are relevant in string theory compactifications and present instead real orbifold singularities.
There is a vast literature on this subject. See {\it e.g.} \cite{Aspinwall} for a review that includes
the study of $\CC^2/\Z_N$ orbifold singularities in a string theory context.}
We explicitly show how the orbifold gauge and Lorentz actions on fermions
are reproduced on ${\cal M}$ by assigning non-trivial gauge and gravitational holonomies
around the four spherical caps, as expected. The gauge holonomies are most easily introduced by switching
on a suitable constant $U(1)$ field strength $F=dA$, whereas the gravitational ones
are automatically the right ones, by construction.
The fluxes reproducing the orbifold $\Z_2$-holonomies are not uniquely determined, whereas
the Atiyah-Singer index theorem predicts that the number of left-handed minus the number
of right-handed fermions on ${\cal M}$ is given by the total amount of flux on ${\cal M}$.

In each region, the massless Dirac equation is straightforward.
It is either trivial (flat space) or analogous to the Dirac equation on a sphere with
a $U(1)$ background $A$, whose solutions are well-known. The non-trivial issue is to find
a globally well-defined wave function on the whole ${\cal M}$. It turns out that it is possible
to find the wave functions for the fermion zero modes in the limit of vanishing spherical
caps, in which case they are given by suitable products and ratios of theta functions.
We always find that all the massless modes have the same two-dimensional chirality, leading
then to 4D fermions of the same chirality and $U(1)$ charge.
Moreover, most of the wave functions get localized
at the fixed points in the orbifold limit, and thus give rise to localized fermion fields.

Various mechanisms for fermion localization in field theory have been studied in the literature.
Fermion localization was originally shown to arise from topological defects in \cite{localization}.
More recently, in a higher dimensional field theory context, localization of fields in both flat and
warped spaces has been considered in \cite{localization-new,localizationFI} (see also \cite{RDaemi}).
The localization phenomenon which we observe here is somehow analogous, as it is
induced by the presence of a background field with
non trivial profile along the extra dimensions. What is particularly interesting in our case
is that this background is precisely the one which is needed for resolving the orbifold singularity.

Depending on the gauge flux, a constant wave function (bulk fermion) can be present or not.
The arbitrariness on the choice of the gauge flux on ${\cal M}$ corresponds to the arbitrariness
of adding states localized at the orbifold fixed points.
However, as already pointed out, these states must all have the same chirality of the bulk fermion
to admit a resolution. Summarizing, we find that one 6D chiral fermion on the resolved space ${\cal M}$
with flux $\kappa\in {\mathbb Z}$ gives rise, in the orbifold limit,
to either 1 bulk and $\kappa-1$ localized fermions, or to $\kappa$ localized fermions.
What option is realized depends on how the resolution is performed, that also fixes
the precise localization pattern of the fermions.
In any case, however, all fermions have the same 4D chirality and $U(1)$
charge.\footnote{At the quantum level, the above theories are anomalous. In a more realistic
situation, one might want to require an anomaly free fermion spectrum.
However, this requirement is not related with the constraints arising from
the orbifold resolution.}
Interestingly enough, we find that sometimes a fermion on ${\cal M}$ gives rise, in the orbifold
limit, to a fermion field that is localized at more than one fixed point, resulting
in a multiple localization of the state.

We also study the limit in which $T^2/\Z_2$ degenerates to $S^1/\Z_2$ and
the resulting resolved space ${\cal M}$ degenerates to a cigar-like surface.
The Dirac equation can be exactly studied in this case, also for finite spherical caps and for
massive fermions.
This study provides a consistency check of the above construction and gives interesting
insights on the structure of the orbifold $S^1/\Z_2$. As straightforward corollary of our result,
all the $S^1/\Z_2$ orbifolds arising as degeneration
limits of a $T^2/\Z_2$ orbifold must have chiral fermions of the same chirality (and $U(1)$ charge)
to admit a ``resolution''.

This restriction on the chirality and $U(1)$ charge of states applies when studying a single fermion
on the resolved space ${\cal M}$. More generally,
one can have $m_i$ fermions with $U(1)$ charges $q_i$
(normalized so that the lowest is set equal to one) and 6D chiralities $\eta_i=\pm 1$ on ${\cal M}$
with flux $\kappa$. In this case one gets, in the orbifold limit,
either $m_i\kappa$ bulk and $(|q_i|-1)m_i\kappa$ localized fermions
with 4D chiralities $\eta_i{\rm{sign}}\ (q_i)$, or
$|q_i|m_i\kappa$ localized fermions with 4D chiralities $\eta_i{\rm{sign}}\ (q_i)$.\footnote{Notice
that all $q_i$ must be integer numbers, since the field strength $F$ acts as a $U(1)$ monopole and
hence a Dirac quantization condition on the fermion charges is required.}
Again, what option is realized depends on how the resolution is performed, as well as
the positions where the fields are localized.
It is clear that these fermion configurations are only a limited class, compared
to the most general choice of bulk and localized fermion fields one could
have on the orbifold.

For a non-abelian theory, the constraint will depend on the gauge twist matrix and it
will still provide a strong constraint on the allowed massless fermion spectra.
As a simple instance of a non-abelian theory, an $SU(2)$ model is briefly discussed in the
final section.

The most important result of this work can be stated as follows:
{\it orbifold field theories can admit a simple resolution, which however implies a strong restriction
on the massless spectrum of fermions one can introduce}. We think that this is an important
constraint that should be taken into account in building models in extra dimensions.
We postpone to a final section some possible lines of development that might arise from this work.

In order to convince the skeptical reader that our results are not an artifact
of the ``cut and paste'' regularization used, we include in our paper a study of the Dirac equation
on a smooth resolution of a conical singularity. This is provided by taking a
family of hyperboloids which, in a suitable limit, degenerates to the cone $\CC/\Z_N$.
We explicitly show that, as far as the massless Dirac equation is concerned, the results
on this space are identical with those obtained by the ``cut and paste'' resolution.

The paper is organized as follows.
In section 2 we illustrate the method we will follow for resolving conical singularities by applying it
to the non-compact cone $\CC/\Z_N$. We solve the massless Dirac equation on the resolved space and
explicitly show how two different resolutions,
the full-fledged smooth one provided by a family of hyperboloids
and the one given by our ``cut and paste'' procedure, gives rise to identical results.

In section 3 we study the Dirac equation on ${\cal M}$ and find how the fermion zero modes
on this space give rise to either bulk or localized fermions in the $T^2/\Z_2$ orbifold limit.
In section 4 we study the limit in which $T^2/\Z_2$ becomes an $S^1/\Z_2$ orbifold.
In this case, the Dirac equation is much easier and can be studied more generally for
finite, and not only infinitesimal, resolutions. As a further consistency check of our
approach, we also study the massive Dirac equation, showing how the usual sine and cosine wave
functions for a bulk fermion on $S^1/\Z_2$ are reproduced.
In section 5 we provide some comments and outline some lines for future research.
Finally, two appendices are included. In the first, we review the study of the spinor zero modes arising
from the compactification on the 2 sphere $S^2$ in presence of a $U(1)$ flux.
In the second, we provide some basic information on the theta functions.

\section{Resolution of  $\CC/\Z_N$ Orbifolds}

\subsection{Definition of the resolving space}

In this section we will consider the non-compact $\CC/\Z_N$ orbifolds which are obtained by the complex
plane $\CC$ with euclidean metric by identifying points connected by a $2\pi/N$ rotation around the origin.
The fundamental domain
of this orbifold is a $2\pi/N$ plane angle, which we choose to have one end on the real positive semi-axis, as
shown in Fig.~1. We consider on this space a $2D$ Dirac fermion field\footnote{In
this and the following sections we will always consider Dirac fermion fields on $2D$ euclidean spaces.
Note however that their chiral components are thought as the wave functions in the internal space of the
$4D$ chiral fermions which arise from a $6D$ chiral fermion when compactified on the two-dimensional
space, as discussed
in Appendix A. Hence the subscripts $R,L$ refer also to $4D$ chirality.}
\be
\psi=\left(\begin{array}{c}\psi_R\\ \psi_L \end{array}\right)\,,
\ee
endowed with a $U(1)$ gauge symmetry under phase transformations.
It is defined as a field on the complex plane $\CC$
which remains invariant under a $2\pi/N$ rotation around the origin, modulo a suitable phase
transformation:
\be
\psi(e^{2\pi i/N} z)={\mc P}\psi(z)\,,\;\;\;\;\;\;\;\;{\mc P}={\ds e^{\pi i\left(1-\frac1N\right){\sigma_3}}
e^{\frac{\pi i}{N}p} }\,,
\label{bcN}
\ee
where $p$ is any integer running from $-N+1$ to $N-1$ at steps of 2.
The $U(1)$ phase in Eq.~(\ref{bcN}) is chosen to make ${\mc P}^N=1$ and an extra $-1=e^{\pi i\sigma_3}$
--- representation on spinors of a $2\pi$ Lorentz rotation ---
has been included in the Lorentz part of ${\mc P}$ for future convenience.

We now define a class of regular spaces which reproduce this orbifold in a suitable limit.
To do so, remove the singularity by cutting from the fundamental domain
a small disk of radius $\epsilon$, as shown in Fig.~1.
 The resulting space, in the $\RR^3$ embedding, is a truncated cone of angle $\alpha$,
such that $\sin{\alpha}=1/N$ (see Fig.~1), which we describe
on the complex plane with a coordinate $z$.
{}From Eq.~(\ref{bcN}) we see that the fermion field $\psi^{(0)}(z)$ in the fundamental domain set
(which we call the ``$(0)$'' set) is
not single-valued on the cone, but there is a \emph{cut} on the line where the two edges
of the fundamental domain glue to produce the truncated cone.
The fermion field is not continuous across the cut.
The value of the field at the final point of the infinitesimal oriented path $C$ crossing the cut shown
in Fig.~1 is related to the one at the initial point by
\be
\psi^{(0)} \left(z_{fin}\right)={\mc P}^{-1}\psi^{(0)} \left(z_{in}\right)\,.
\label{cutN}
\ee
This cut is a well defined object. To be precise, we should remember that the
(truncated) fundamental domain of the orbifold is not a good chart to describe all the points of
the (truncated) cone. The two edges of the fundamental domain, indeed, do not really overlap but they are
separated by an infinitesimal distance. The remaining ``line'' on the cone which is not covered by the
fundamental domain is covered by a second, infinitesimally thin chart which intersects with the fundamental
domain chart on both sides of the cut. The transition functions on the two sides can of course be different
and this makes a discontinuity for the field in the $(0)$ chart as in Eq.~(\ref{cutN}).
The $1$-dimensional analogue of this situation is the $S^1$ circle with twisted boundary conditions
\`a la Scherk-Schwarz.

\begin{figure}[t]
\begin{center}
\begin{picture}(300,150)
\psfrag{g}{\footnotesize $\gamma$}
\psfrag{a}{\footnotesize $\alpha$}
\psfrag{e}{\footnotesize $\epsilon$}
\psfrag{s}{\footnotesize $\epsilon /N$}
\psfrag{c}{\footnotesize $C$}
\psfrag{p}{\footnotesize $2\pi/N$}
\put(-80,20){\includegraphics*[width=8cm]{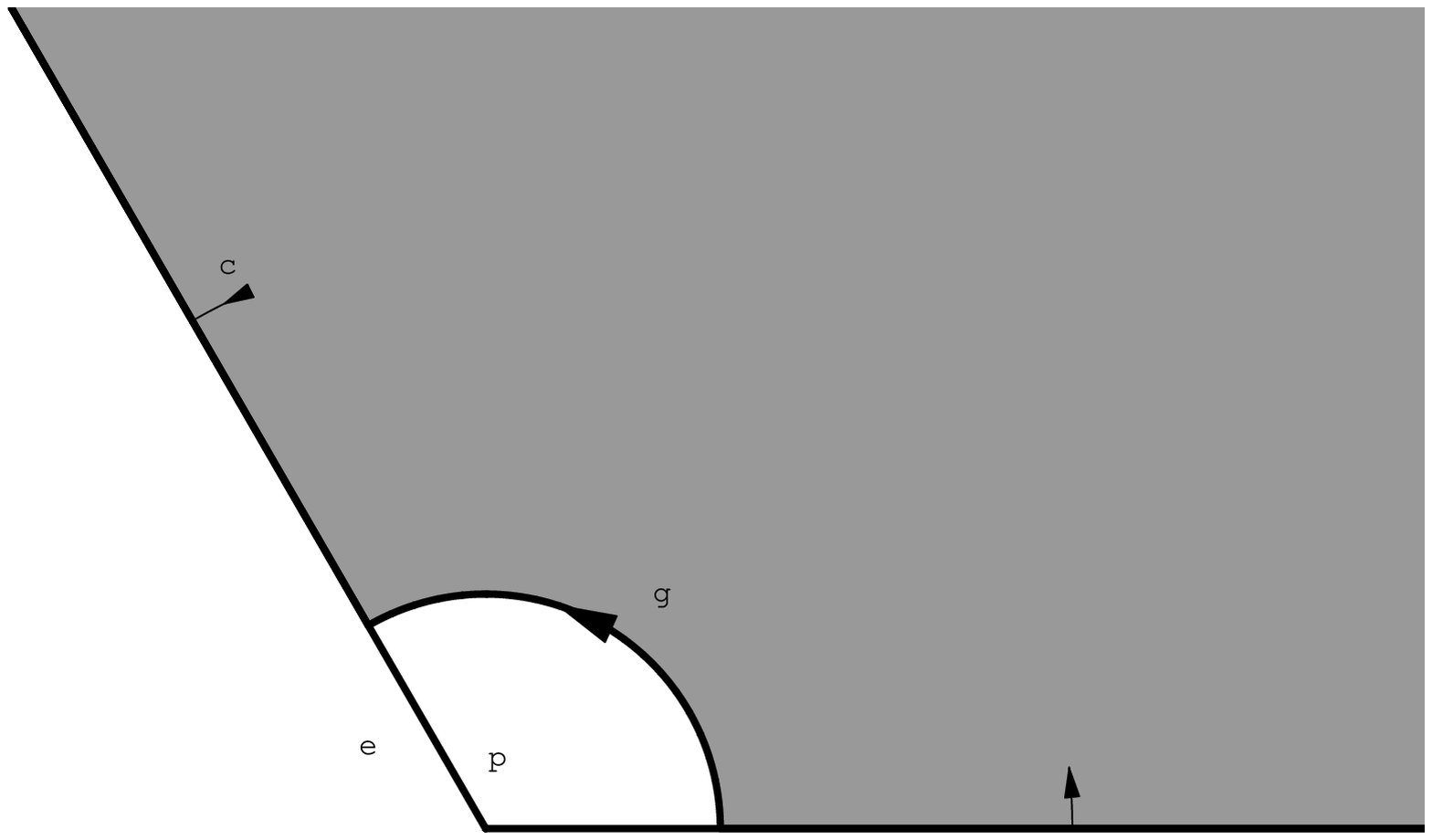}}
\put(200,0){\includegraphics*[width=5cm]{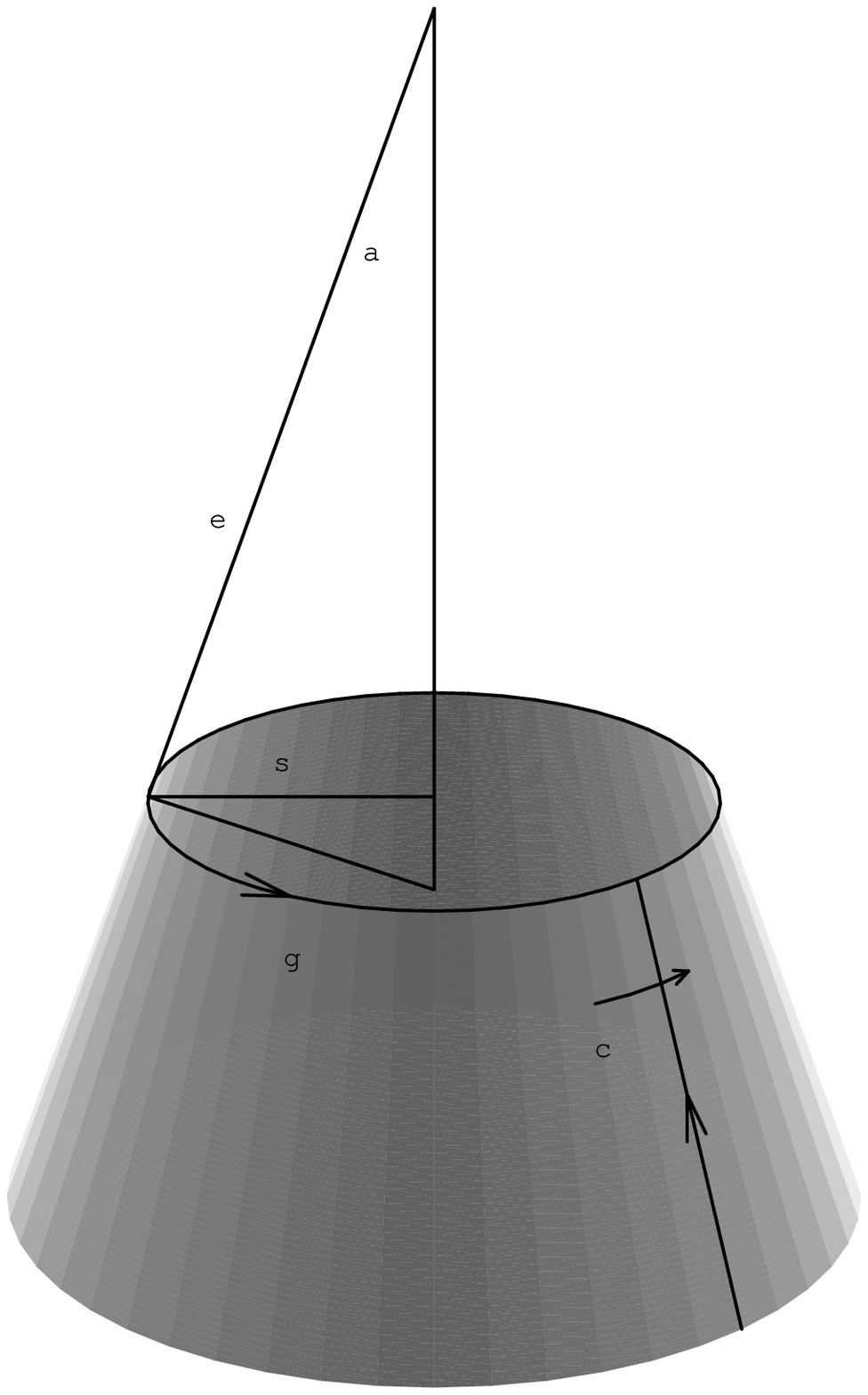}}
\end{picture}
\caption{\footnotesize{The picture in the left shows the fundamental domain of the $\CC/\Z_3$ orbifold from
which a disk of radius $\epsilon$ has been removed.
The resulting truncated cone, including the circle $\gamma$ and the
cut which runs along the cone, is shown in the right.
The infinitesimal path $C$, oriented in such a way that the vector product of
its direction with the one of the cut is outgoing from the surface, is also shown.}}
\end{center}
\protect\label{fig.1}
\end{figure}
The truncated orbifold we are describing is a well defined regular space with zero gauge and spin
connection everywhere, the only non trivial structure being provided by the cut. We want now to complete
this space by attaching a suitable surface to the oriented closed curve $\gamma$, shown in Fig.~1,
which is a boundary for our truncated orbifold. Note that $\gamma$ is a circle of radius
$\epsilon/N$, so it shrinks to a point in the orbifold limit $\epsilon\rightarrow0$.

The new surface to be attached to $\gamma$ will be described by a set which we call ``$(1)$'' and neither
the gauge nor the spin
connections, $A_1$ and $\omega_1$ respectively, will be trivial on it. Indeed, due to the presence of the
cut, a spinor field parallel transported along the closed curve $\gamma$ is rotated by an amount
\be
\psi_{1-turn}=W\psi\,,
\label{hol}
\ee
where $W={\mc P}^{-1}$ is the holonomy of $\gamma$. In the (0) set the gauge and spin connections
vanish and this holonomy is provided by the cut. In the
$(1)$ set no cuts have to be present and the same holonomy (Eq.~(\ref{hol}) is of course gauge covariant)
must be provided by a Wilson line:
\be
{\ds e^{-i\oint_{\gamma}\left(A_1+\frac12\sigma_3\omega_1\right)}}
={\ds e^{-i\int_{1}\left(F+\frac12\sigma_3R\right)}}= W=
{\mc P}^{-1}={\ds e^{-\pi i\left(1-\frac1N\right){\sigma_3}}e^{-\frac{\pi i}{N}p}} \,,
\label{Wl}
\ee
where $F$ and $R$ are the field strength and curvature two-forms, Stokes' theorem has been used
and unit charge to $\psi$ under the $U(1)$ local symmetry has been assigned.
We see that Eq.~(\ref{Wl}) only gives a quantization condition on the integrals
over the $(1)$ set of the gauge and gravitational curvatures, it does not fix them completely. One set of
solutions to Eq.~(\ref{Wl}) is provided by\footnote{The reason why we focus on this class of solutions
 will be clear in the following; note that it is not the generic solution to Eq.~(\ref{Wl}).
In particular, we could modify the first line of Eq.~(\ref{qc}) by adding integers to the r.h.s.; this should
correspond to adding handles to the regularizing space.}
\be
\frac1{4\pi}\int_1 R=\frac12\left(1-\frac1N\right)\,,\ \ \ \ \
\frac1{2\pi}\int_1 F=\frac{p}{2N}+n\,,
\label{qc}
\ee
where $n$ is an arbitrary integer. We will see that the ambiguity in the choice of the integer $n$
in the second expression of Eq.~(\ref{qc}) is not at all irrelevant.
The flux for $F$ is indeed related to the index theorem
in two dimensions and then, when repeating the above discussion for each of the conical singularities
of the compact $T^2/\Z_N$ orbifolds (we will see the case of $T^2/\Z_2$ explicitly), this ambiguity will affect
the index of the Dirac operator, and then the number of massless states.

Summarizing, in this subsection we have discussed a general class of resolutions of the $\CC/\Z_N$ orbifold,
obtained by gluing a flat truncated cone with a cut along it,
to a non flat surface which substitutes the singularity and shrinks to a point in the orbifold limit.
It has been shown how the integral of the gauge and gravitational curvatures on this surface are quantized,
but not completely fixed. We will now discuss an explicit example of such spaces, in which the surface which
parametrizes the resolved singularity is a spherical cap. We will study the massless Dirac equation on this space and show
that, as expected, changing the value of the field-strength as allowed by Eq.~(\ref{qc}) corresponds to changing the
number of chiral zero modes.

\subsection{Resolution through a spherical cap}

Consider the truncated cone in Fig.~1. We want to glue a spherical cap
to the circle $\gamma$ to complete the regularizing manifold. Our intuition suggests us to use
a suitable portion of sphere so that the tangent space is well defined on $\gamma$.
\begin{figure}[t]
\begin{center}
\begin{picture}(200,200)
\psfrag{g}{\footnotesize $\gamma$}
\psfrag{a}{\footnotesize $\alpha$}
\psfrag{e}{\footnotesize $\epsilon$}
\psfrag{s}{\footnotesize $\epsilon /N$}
\psfrag{c}{\footnotesize $C$}
\psfrag{p}{\footnotesize $2\pi/N$}
\put(-20,0){\includegraphics*[width=8cm]{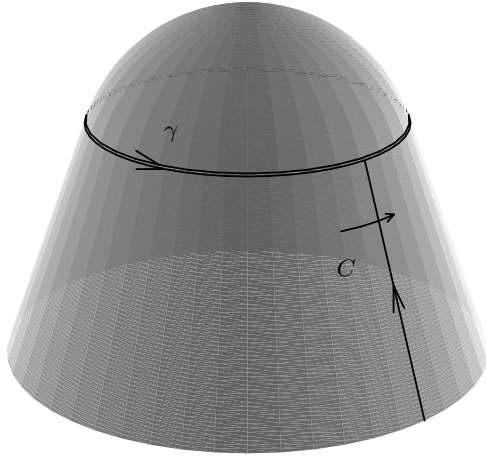}}
\end{picture}
\caption{\footnotesize{The truncated cone corresponding to the $\CC/\Z_3$ orbifold, with
a spherical cap attached. The latter is chosen such that the tangent plane is well defined on $\gamma$.}}
\end{center}
\protect\label{fig.2}
\end{figure}
The resulting space is shown in Fig.~2 and is described by two disjoint sets,
each with its local coordinate system;
on the spherical cap (which we call the $(1)$ set, to follow the notation of the previous subsection) we can use spherical
coordinates with $\theta\in [0,\theta_{Max}=\pi/2-\alpha]\,,\;\phi\in [0,2\pi)$.
On the truncated cone (the $(0)$ set), which is flat, we can use complex coordinates $z$ and $\bar z$.
On the spherical cap, the spin connection $\omega_1$ is the one on $S^2$, namely
$\omega_1 = (1-\cos \theta)d\phi$ as discussed in Appendix A, and then the integral of $R=d\omega_1$ on
the spherical cap
\be
\frac1{4\pi}\int_{1}{R}=\frac12\left(1-\cos{\theta_{Max}}\right)=\frac12\left(1-\frac1N\right)
\ee
matches the result in Eq.~(\ref{qc}) that we derived on general grounds.

As discussed in the previous subsection, a non trivial gauge connection $A_{1}$ must be present
on the spherical cap. We take it to be proportional to $\omega_1$, so that $A_{1}=\kappa/2 \omega_{1}$, with
\be
\kappa=\frac{p+2Nn}{N-1}\,,
\label{kap}
\ee
in order to satisfy the quantization condition given in Eq.~(\ref{qc}).
On the cone, $\omega_{0}=A_{0}=0$.

The massless Dirac equation on this space can now be solved.
Since on the curve $\gamma$ which separates the spherical cap from the truncated cone we have
\be
{\omega_{0}}_{|_\gamma}=\omega_{1}(\theta=\theta_{Max})-(1-\sin{\alpha})d\phi\,,
\ee
and similarly for $A$, the wave function on the spherical cap $\psi_1$ and that on the cone $\psi_0$
are related by a gauge$+$Lorentz transformation:
\be
{\psi_{0}}_{|_\gamma}=\ds{e^{\frac{i}2
\phi(1-\sin{\alpha})(\kappa+\sigma_3)}}\psi_{1}(\theta=\theta_{Max},\;\phi)\,.
\ee
More precisely (see Fig.~1), one has
\be
\psi_{0}(z=\epsilon e^{i\sin{\alpha}\phi})=\ds{e^{\frac{i}2 \phi(1-\sin{\alpha})(\kappa+\sigma_3)}}
\psi_{1}(\theta=\theta_{Max},\;\phi)\,.
\label{gamma-cond}
\ee
The solutions of the Dirac equation on the spherical cap are given by Eqs.(\ref{spheresolu}).
On the other hand, the Dirac equation on the cone simply states that $\psi_{0,R}$ and $\psi_{0,L}$ have to be
generic holomorphic and anti-holomorphic functions, respectively. They will match with the solutions
(\ref{spheresolu}) with definite angular momenta $N_{L,R}$ around $\phi$ only if we take
these functions to be simple monomials, namely
\bea
\psi_{0,L}= \a\a \bar z^{\lambda_L}\,,\nn \\
\psi_{0,R}=\a\a z^{\lambda_R}\,.
\label{sol-flat}
\eea
The constants $\lambda_{L}$ and $\lambda_R$ are determined by means of Eq.~(\ref{gamma-cond}):
\bea
\lambda_L = \a\a \Bigg(1-\frac{1}{\sin{\alpha}}\Bigg)\frac{\kappa -1}2-\frac{N_L}{\sin{\alpha}}
=-\frac12(p-N+1)-N(n+N_L)
 \,,\nn\\
\lambda_R =\a\a \Bigg(\frac{1}{\sin{\alpha}}-1\Bigg)\frac{\kappa +1}2+\frac{N_R}{\sin{\alpha}}
=\frac12(p+N-1)+N(n+N_R)\,,
\label{value-lambda}
\eea
where the angular momenta $N_{L,R}$ are subject to the conditions (\ref{ang-cond}).

If we do not impose normalizability to our wave-functions,
there are an infinite number of solutions to the Dirac equation, depending on the values of
$\lambda_L$ and $\lambda_R$. If we consider those which do not diverge at infinity, one has
\bea
N_L \ge (\sin\alpha-1)\frac{\kappa-1}2=-\frac1{2N}(p-N+1)-n\,,\nn\\
N_R \le (\sin\alpha-1)\frac{\kappa+1}2=-\frac1{2N}(p+N-1)-n\,.
\label{limit}
\eea
When $N_{L,R}$ reach the bounds of the above inequalities, in the orbifold limit $\epsilon\rightarrow0$,
the corresponding wave function is constant on the cone.
Constant solutions are peculiar as they correspond to the usual constant zero modes of the $\CC/\Z_N$ orbifold; they
do not always arise depending on the value of $p$, as expected from Eq.~(\ref{bcN}), but also on the sign of $n$.
The bounds of the inequalities in Eq.~(\ref{limit}) can be reached if
$p=N-1$ and $n\ge0$ for left-handed states or $p=-N+1$ and $n\le0$ for right-handed states.
These values of $p$ are precisely
those required for the orbifold projection given in Eq.~(\ref{bcN}) to leave untwisted the left- and right-handed states
respectively, in such a way that the corresponding zero mode is present.

The non-constant solutions are those for which the strict inequalities are satisfied in Eq.~(\ref{limit}).
For $n>0$, $n$ left-handed states of this kind are present, and no right-handed ones. For $n<0$, we have $-n$
right-handed states and no left-handed ones. For $n=0$, no non-constant state can be present.

We see that, as expected, different choices of the integer $n$, which was not fixed by the general discussion
of the previous subsection,  correspond to a different number of (non-divergent at infinity)
zero modes on the resolution of the orbifold.
What is important to notice is
that the extra states which are introduced by the presence of the integer $n$ have a power-like divergence
on the intersection between the spherical cap and the truncated cone when the orbifold limit $\epsilon\rightarrow 0$
is taken. They correspond then to states which are \emph{localized} at the fixed point of the orbifold which,
surprisingly enough, \emph{naturally} arise when resolving the singularity.

\subsection{A smooth resolution of $\CC/\Z_N$}

The results of the previous subsections suggest us that the ambiguity in the choice of the field strength
localized around the singularities, which arise in Eq.~(\ref{qc}), should correspond to the possibility of
adding massless states localized at the fixed points of the orbifold. Namely, two resolutions
which differ by the choices of the gauge fluxes near the singularities should both resemble the
same orbifold, but with different distribution of bulk and brane fields.

Before verifying the above general statement
in the case of the $T^2/\Z_2$ orbifold, we want to check the validity of the results we derived up to now,
which are affected by a little technical problem. The space we considered in the previous subsection
is indeed not at all a resolution (we will denote it a ``${\cal C}^1$ resolution'')
because the gauge and Lorentz curvatures are discontinuous along the curve
$\gamma$, jumping from zero on the cone to a constant value on the spherical cap. As a consequence, all our discussion
is mathematically not very precise\footnote{Note that, for instance, we thought the intersection of the $(0)$
set with the $(1)$ set to be a line instead of an open set, as it should.} even if every step could be better justified
by thinking to an underlying smooth space that approximates our ${\cal C}^1$ resolution in a suitable limit.

In order to verify that our procedure is anyhow correct, we now consider a smooth ${\cal C}^\infty$
resolution of the $\CC/\Z_N$ orbifold, which is
provided by a suitable surface of a $3D$ hyperboloid, and repeat the study of the massless Dirac equation on this space.
We will show that, in the orbifold limit, the number and profile of the chiral zero modes are exactly
the same as those we found with the simpler ${\cal C}^1$ resolution.
\begin{figure}[t]
\begin{center}
\begin{picture}(200,200)
\psfrag{g}{\footnotesize $\gamma$}
\psfrag{a}{\footnotesize $\alpha$}
\psfrag{e}{\footnotesize $\epsilon$}
\psfrag{s}{\footnotesize $\epsilon /N$}
\psfrag{c}{\footnotesize $C$}
\psfrag{p}{\footnotesize $2\pi/N$}
\put(0,0){\includegraphics*[width=6cm]{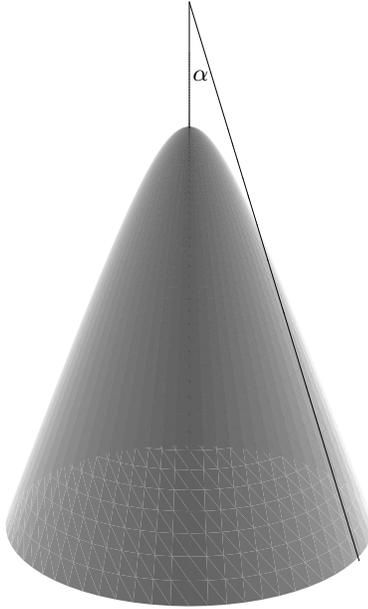}}
\end{picture}
\caption{\footnotesize{The $\RR^3$ embedding of the hyperboloid which resolves the $\CC/\Z_3$ orbifold
is shown.
}}
\end{center}
\protect\label{fig.3}
\end{figure}

As already discussed, the $\CC/\Z_N$ orbifold corresponds
to a cone with an angle $\alpha$ such that $\sin{\alpha}=1/N$. To build a smooth resolution of this orbifold,
consider an hyperboloid in $\RR^3$ such that the angle of its asymptotic cone is $\alpha$, as
shown in Fig.~3. If we want the axis of the asymptotic cone to be the $z$-axis and its vertex
to be the origin, its equation in $\RR^3$ is
\be
\frac{z^2}{\cos^2\alpha}-\frac{(x^2+y^2)}{\sin^2\alpha}=\epsilon^2\,,
\label{h-e}
\ee
with $\epsilon$ an arbitrary real parameter.
Convenient coordinates for the hyperboloid are  $t\in [0,\infty)$ and $\phi\in [0,2\pi)$, in terms of which we can
parametrize its surface as
\be
\left\{\begin{array}{l} x=\epsilon\sin\alpha\sinh{t}\cos{\phi}\\ y=\epsilon\sin\alpha \sinh{t}\sin{\phi} \\
z=\epsilon\cos\alpha\cosh{t}\end{array}\right.\,.
\label{changecoord}
\ee
The induced metric in the (Euclidean) $\RR^3$ embedding reads
\be
ds^2=\epsilon^2\left[\sin^2\alpha\cosh^2{t}+\cos^2\alpha\sinh^2{t}\right]dt^2+
\epsilon^2\sin^2\alpha\sinh^2{t}d\phi^2\,,
\label{conemetric}
\ee
and the corresponding spin connection which is well defined everywhere is given by
\be
\omega=\left[1-\frac{\sin\alpha\cosh{t}}{\sqrt{\sin^2\alpha\cosh^2{t}+\cos^2\alpha\sinh^2{t}}}\right] d\phi\,.
\label{sch}
\ee

As we will verify later, in order for this space to resemble the $\CC/\Z_N$ orbifold with a projection matrix as in
Eq.~(\ref{bcN}), it must be endowed with a non-trivial gauge connection $A=\kappa/2\omega$ with $\kappa$ as in
Eq.~(\ref{kap}). We see from Eqs.~(\ref{h-e}) and (\ref{changecoord}) that this space reproduces the cone
 in the limit  $\epsilon\rightarrow0$ and that, in this limit,
one has to consider $t\rightarrow \infty$ with $\rho = \epsilon e^t/2$ fixed. It is also clear that $(\rho, \phi)$
will be the polar coordinates of the cone, related to the complex coordinate $z$ that we used in the previous
subsection as $z=\rho e^{i\phi/N}$.
In the orbifold limit, the metric (\ref{conemetric}) becomes
\be
ds^2 \;\rightarrow\; d\rho^2 + \rho^2 \sin\alpha^2 d\phi^2=dzd\ov{z} \,,
\ee
that is, indeed, the metric of a cone with angle $\alpha$. In the same limit,
the spin connection in Eq.~(\ref{sch}) reads
\be
\omega\;\rightarrow\;\left(1-\sin\alpha\right)d\phi=\left(1-\frac1N\right)d\phi\,,
\label{ol}
\ee
and, correspondingly, one has for the gauge connection $A=\kappa/2\omega$:
\be
A\;\rightarrow\;\left(\frac{p}{2N}+n\right)d\phi\,,
\label{al}
\ee
having substituted the value of $\kappa$ given in Eq.~(\ref{kap}).

Apparently, the cone which arises from this hyperboloid does not resemble, in the orbifold limit and
away from the singularity, the truncated cone which corresponds to the $\CC/\Z_N$ orbifold.
In the former, there are no cuts and non-vanishing gauge and spin connections; in the latter
there is a cut and vanishing gauge and spin connections. The two situations are however
completely equivalent as they are related by a gauge$+$Lorentz transformation. In the orbifold limit,
indeed, the holonomy of any circuit surrounding the singularity, as computed from Eqs.~(\ref{ol}) and (\ref{al}),
exactly matches the one in Eq.~(\ref{Wl}).
We now consider the massless Dirac equation for a two-dimensional Dirac spinor
$\psi=(\psi_R,\psi_L)^t$ on the hyperboloid. It reads
\be
\left\{\begin{array}{c} \ds{\frac{1}{\epsilon\sqrt{\sin^2\alpha\cosh^2{t}+\cos^2\alpha\sinh^2{t}}}
\partial_t\psi_L-
\frac{i}{\epsilon\sin\alpha\sinh{t}}\left[\partial_\phi +\frac{i}2(k-1)\omega_\phi \right]\psi_L=0}\\
\ds{\frac{1}{\epsilon\sqrt{\sin^2\alpha\cosh^2{t}+\cos^2\alpha\sinh^2{t}}}\partial_t\psi_R+
\frac{i}{\epsilon\sin\alpha\sinh{t}}\left[\partial_\phi +\frac{i}2(k+1)\omega_\phi \right]\psi_R=0}
 \end{array}\right.\,.
\label{DiracHyperI}
\ee
If we take
\be
\psi_{L,R}=f_{L,R}(t)\ds{e^{iN_{L,R}\phi}}\,,
\label{pf}
\ee
with $N_{L,R}$ integers so that the wave functions are single valued,
Eq.~(\ref{DiracHyperI}) gives
\be
\left\{ \begin{array}{l} \partial_t{\log{f_L}}=\frac12(k-1)\coth{t}-(N_L+\frac{k-1}2)
\frac{\sqrt{\sin^2\alpha\cosh^2{t}+\cos^2\alpha\sinh^2{t}}}{\sin\alpha\sinh{t}}\\
\partial_t{\log{f_R}}=-\frac12(k+1)\coth{t}+(N_R+\frac{k+1}2)
\frac{\sqrt{\sin^2\alpha\cosh^2{t}+\cos^2\alpha\sinh^2{t}}}{\sin\alpha\sinh{t}}
 \end{array}\right.\,.
\label{DiracHyperII}
\ee
The solutions of the above equations could be explicitly found, but we will only need their limiting behaviors
as $t\rightarrow 0$ and $t\rightarrow \infty$.
In the limit $t\sim 0$, the solutions behave as
\be
\left\{\begin{array}{l} f_L\sim \ds{t^{-N_L}}\\f_R\sim \ds{t^{+N_R}} \end{array}\right.\,.
\ee
Requiring them to be single-valued at $t=0$ implies $N_L\leq 0$, $N_R\geq 0$,
as it happens for fermions on $S^2$ (see the Appendix A). At $t\sim\infty$, {\it i.e.} in the orbifold limit,
the behavior is
\be
\left\{\begin{array}{c} f_L\sim \ds{e^{\left[\frac{k-1}2\left(1-\frac1{\sin{\alpha}}\right)-
\frac{N_L}{\sin{\alpha}}\right]t}}\\
f_R\sim \ds{e^{\left[\frac{k+1}2\left(-1+\frac1{\sin{\alpha}}\right)+\frac{N_R}{\sin{\alpha}}\right]t}}
\end{array}\right.\,.
\label{DiracHyperIII}
\ee
Once expressed in terms of the coordinate $|z|=\rho=\epsilon e^t/2$, Eq.~(\ref{DiracHyperIII})
precisely matches the power-like behavior we found
in Eqs.~(\ref{sol-flat}) and (\ref{value-lambda}) for the wave functions on the truncated cone
arising from the ${\cal C}^1$ resolution with the spherical cap.
A solution to the massless Dirac equation on the hyperboloid
with a certain angular momentum $N_{L,R}$ corresponds then to the one arising from the ${\cal C}^1$ resolution with the same angular
momentum. The reader can verify, making use of the relation $z=\rho e^{i\phi/N}$ to compare Eq.~(\ref{pf}) with
Eq.~(\ref{sol-flat}), that the $\phi$ behavior of the solutions is also consistent with this identification,
once the gauge$+$Lorentz transformation which is required to make the cut on the truncated cone disappear is performed.

Summarizing, the behavior in the orbifold limit of the wave functions of chiral fermions found
in Sect.~2.2 with the ${\cal C}^1$ resolution is not an artifact of our ``cut and paste'' method,
as these functions are also reproduced when a more precise resolution is performed.
This confirms all the results on fermion localization that we discussed
in Sect.~2.2 and gives us confidence on the validity of the ${\cal C}^1$ resolution method that
we will follow in the next section to resolve the $T^2/\Z_2$ orbifold.

\section{The ${\cal C}^1$ resolution of the orbifold $T^2/\Z_2$}

The orbifold $T^2/\Z_2$ is defined by identifying the points on $T^2$ which are related by
the $\pi$ Lorentz rotation $z\rightarrow -z$. The covering torus $T^2$ is defined, as usual, by
identifying points in the complex plane as
\be
z\sim z \, + \, m\, +\, n\, \tau\,,
\label{lattice}
\ee
where $\tau$ denotes the complex structure of $T^2$ and $m,n$ are arbitrary integers.
A fixed point $z_i$ on $T^2/\Z_2$ satisfies the relation
\be
z_i = -z_i + m_i + n_i\tau \,,
\label{fixed}
\ee
with $m_i,n_i$ arbitrary integers.
Eq.~(\ref{fixed}) has
four independent solutions, namely $z_1=0$ ($m_1=n_1=0$), $z_2=1/2$ ($m_2=1,n_2=0$),
$z_3=(1+\tau)/2$ ($m_3=n_3=1$), $z_4=\tau/2$ ($m_4=0,n_4=1$), which are the four fixed points of
$T^2/\Z_2$. At each of the four fixed points, the space has a conical $\CC/\Z_2$ singularity.

Consider a $2D$ Dirac fermion $\psi$ on this space, of unit charge under a $U(1)$ local
symmetry. It is defined as a Dirac field on the torus which remains invariant under $\pi$ rotations,
modulo a suitable $U(1)$ transformation:
\be
\psi(-z)=\mc{P}\,\psi(z)\,,
\label{or-cond}
\ee
where
\be
\mc{P}=\ds{ e^{\frac{i\pi}{2}(\sigma_3+p)}}\,,
\ee
with $p$ either $+1$ or $-1$, according to Eq.~(\ref{bcN}).
Moreover, $\psi(z)$ must be a field on the torus and then satisfies the conditions
\bea
\psi(z+1)= \a\a T_1\psi(z)\,, \nn \\
\psi(z+\tau)= \a\a T_2\psi(z)\,,
\label{per-cond}
\eea
where
\be
T_{1}=\ds{e^{i\pi t_{1}}}\,,\;\;\;\;T_{2}=\ds{e^{-i\pi t_{2}}}\,,
\label{T-6D}
\ee
and $t_{1,2}$ either $0$ or $1$.
The integers $p$, $t_1$ and $t_2$ denote the gauge action on $\psi$: in particular, they
represent how the $\Z_2$ action and the two translations defining the orbifold are
embedded in the $U(1)$ gauge group. On a simply connected space such as $T^2$,
$T_1$ and $T_2$ denote the two independent Wilson lines one can have around
the two cycles of the torus, and can take any value. On $T^2/\Z_2$, consistency with
the parity projection fixes $T_{1,2}$ to the discrete values $\pm 1$.
If $t_1$ or $t_2$ are different from zero, there is no bulk massless fermion,
whereas if they both vanish we have a left-handed or right-handed fermion depending
on whether $p=+1$ or $p=-1$, respectively.

\begin{figure}[t]
\begin{center}
\begin{picture}(200,200)
\psfrag{g1}{\footnotesize $\gamma_1$}
\psfrag{g2}{\footnotesize $\gamma_2$}
\psfrag{g3}{\footnotesize $\gamma_3$}
\psfrag{g4}{\footnotesize $\gamma_4$}
\psfrag{g1p}{\footnotesize $\gamma_{1}'$}
\psfrag{g2p}{\footnotesize $\gamma_{2}'$}
\psfrag{A}{\footnotesize $A$}
\psfrag{B}{\footnotesize $B$}
\psfrag{C}{\footnotesize $C$}
\psfrag{1}{\footnotesize $1$}
\psfrag{2}{\footnotesize $2$}
\psfrag{3}{\footnotesize $3$}
\psfrag{4}{\footnotesize $4$}
\put(-20,0){\includegraphics*[width=10cm]{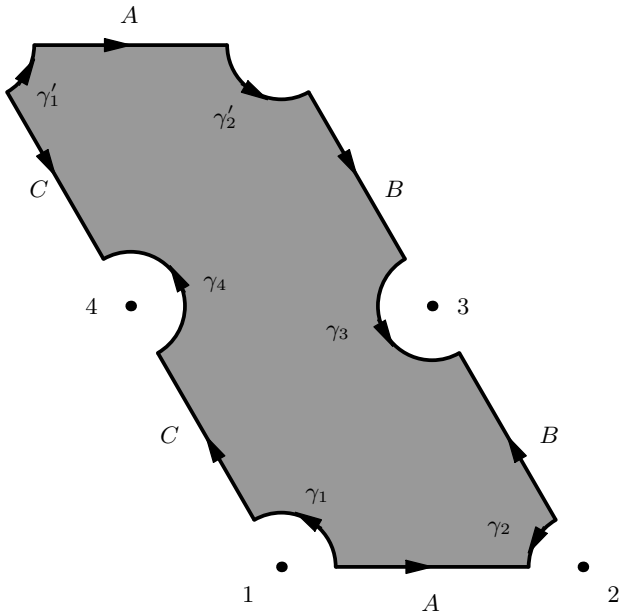}}
\end{picture}
\caption{\footnotesize{The fundamental domain in the complex plane of the $T^2/\Z_2$
orbifold with complex structure $\tau=e^{2\pi i/3}$. The location of the four fixed points, labeled as defined in
the text below Eq.~(\ref{fixed}), is also shown. The fixed points have been removed by cutting infinitesimal
disks of radius $\epsilon$
away from the fundamental domain. Its edges are given by the six oriented lines shown in the figure.
Each couple of lines is identified due to torus$+$orbifold point identifications.
Due to these identifications, the oriented curves $\gamma_3$, $\gamma_4$, $\gamma_1+\gamma_{1}'$
and $\gamma_2+\gamma_{2}'$ are closed and represent the boundaries of the ``truncated'' orbifold.
}}
\end{center}
\protect\label{fig.4}
\end{figure}

In addition to bulk fermions, in an orbifold field theory
one can generally add an arbitrary number $m_i$, $i=1,\ldots4$,
of chiral fermions $\chi_i$ localized at each fixed point $z_i$,
whose chirality and $U(1)$ charge is arbitrary. Our main aim is to understand whether this general set-up
of a bulk fermion and an arbitrary number of localized fermions admits a resolution.

The ${\cal C}^1$ resolution of the $T^2/\Z_2$ orbifold can be built as follows.
As already mentioned, this space --- whose
fundamental domain in the complex plane is shown in Fig.~4 --- has four conical singularities of the form
$\CC/\Z_2$ at the four fixed points. By means of Eqs.~(\ref{fixed})--(\ref{per-cond}) and looking at Fig.~4,
the holonomies of the four infinitesimal circuits $\gamma_i$ surrounding the singularities can be computed.
One can verify that the four conical singularities of the orbifold are equivalent to the ones of
four $\CC/\Z_2$ orbifolds with projection matrices $\mc{P}_i$
given by
\be
\mc{P}_1 = \mc{P}\,, \;\;\; \mc{P}_2 = T_1 \mc{P}\,, \;\;\; \mc{P}_3 = T_1 T_2 \mc{P}\,, \;\;\; \mc{P}_4 = T_2 \mc{P}\,,
\label{Pi}
\ee
which are indeed the effective projections around each fixed point $z_i$, defined as the transformation matrix
of the field under a $\pi$ rotation around the $i$-th fixed point: $\psi(-z+2z_i)\equiv\mc{P}_i\psi(z)$.
The $T^2/\Z_2$ orbifold can be visualized \cite{Heb} as two rectangles glued at the edges. The presence of
$3$ non trivial cuts along the $4$ edges (one cut can be always gauged away) motivates the presence of
non trivial holonomies around the fixed points. We see from Eq.~(\ref{Pi}) that the $4$ holonomies are indeed
expressed as products of $3$ matrices only: $T_{1,2}$ and $\mc{P}$.

By applying the general results of Sect.2 to the case $N=2$, we see that a ${\cal C}^1$ resolution of the
$T^2/\Z_2$ orbifold is obtained by attaching to the four circles $\gamma_i$, four spherical caps
(we will call them the ``(i)'' sets) with
coordinates $\theta_i\in [0,\pi/3]\,,\;\phi_i\in [0,2\pi)$. On each of them a non-trivial gauge
background is present:
\bea
A_{i}= \a\a \frac{k_i}{2}\,\omega_{i}\,, \nn \\
\omega_i = \a\a (1-\cos \theta_i) \, d\phi_i\,,\nn
\eea
with $\kappa_i$ of the form
\be
\kappa_i=p_i+4n_i\,,
\label{ki}
\ee
where $n_i$ are arbitrary integers and $p_i$ are either $+1$ or $-1$,
depending on the values of $p,\,t_1$ and $t_2$. Namely, the $p_i$'s are defined as
\be
\begin{array}{l}
p_1=p\,,\\p_2=(p+2t_1)\;{\rm Mod}\,4=(-)^{t_1}p\,,\\
p_3=(p+2t_1-2t_2)\;{\rm Mod}\,4=(-)^{t_1+t_2}p\,,\\
p_4=(p-2t_2)\;{\rm Mod}\,4=(-)^{t_2}p\,,
\end{array}
\label{pi}
\ee
as implied by Eq.~(\ref{Pi}).

The manifold is described by five sets: the four spherical caps
and the remaining flat region connecting them. We denote the flat region as ``(0)'' set and we take
vanishing gauge and spin connections $A_0$ and $\omega_0$ on it.

As already discussed in Sect.2, the values of the fluxes on each spherical cap are not uniquely
determined by the orbifold periodicity conditions, that do not depend on $n_i$. On the contrary, the
Atiyah-Singer index theorem applied to this space predicts that
\be
n_L-n_R = \frac{1}{2\pi} \int_{{\cal M}} F = \frac{1}{2\pi} \sum_{i=1}^4 \int_{i} F_i =
\frac 14 \sum_{i=1}^4 \kappa_i
\equiv \kappa \,,
\label{AS-index}
\ee
where $n_{L(R)}$ are the number of left (right) handed massless fermions on the resolved orbifold ${\cal M}$.
Note that $\kappa$, as defined in Eq.~(\ref{AS-index}),
can be rewritten as
\be
\kappa=\frac14\left[1+(-)^{t_1}\right]\left[1+(-)^{t_2}\right]p+\sum_{i=1}^4n_i\,,
\ee
and is therefore an integer number as it should, being the index of the Dirac operator.
We will see in next subsection how these chiral fermions predicted by the index theorem,
whose number depend on the integers $n_i$, arise and how they are related to the orbifold theory.

\subsection{Dirac equation}

On the five sets composing ${\cal M}$, the massless Dirac equation can easily be solved.
In each spherical cap, the solution $\psi_i$ is a generic linear combination of the ones
in Eq.~(\ref{spheresolu}) with $\kappa=\kappa_i$;
the latter are labeled by the angular momenta $N_{L,R}^{(i)}$ which
must have definite sign as in Eq.~(\ref{ang-cond}).
In the intermediate flat region, as for the case of the cone discussed before,
the Dirac equation implies that $\psi_{0,R}$ and $\psi_{0,L}$ have to be
holomorphic and anti-holomorphic functions, respectively, satisfying the periodicity
conditions (\ref{per-cond}) along the two cycles of the covering torus.\footnote{Indeed, the local resolution
of the singularities that we have made does not change the global periodicity of the
wave functions.} The non-trivial issue is to find which holomorphic and anti-holomorphic functions
$\psi_{0,R}$ and $\psi_{0,L}$ match with the wave functions $\psi_i$ to give a properly
well-defined wave function on ${\cal M}$.
In order to do that, we have to connect $\omega_0$ and $\omega_i$ in the intersection
curves $\gamma_i$, whose points are described by $(\theta_i=\pi/3,\,\phi_i)$ ($\phi_i\in [0,2\pi)$) in the $(i)$
coordinate system.
This gives
\be
\left\{\begin{array}{c} {\ds {\omega_{0}}_{|_{\gamma_i}}}=\omega_{i}(\theta_i=\pi/3,\,\phi_i)-\frac12 d\phi_{i} \\
{A_{0}}_{|_{\gamma_i}}=A_{i}(\theta_i=\pi/3\,,\phi_i)-\frac{\kappa_i}4 d\phi_{i}
\end{array}\right.  \Rightarrow\;\;\;\;
{\psi_{0}}_{|_{\gamma_i}}=\ds{e^{\frac{i}4 \phi_{i}(\kappa_i+\sigma_3)}}\psi_{i}(\theta_i=\pi/3\,,\phi_i)\,.
\label{sol-hyp}
\ee
More explicitly (see Fig.~4)
\be
\psi_{0}\left(z=z_i+\epsilon e^{\frac{i}2\phi_i}\right)=
\ds{e^{\frac{i}4 \phi_{i}(\kappa_i+\sigma_3)}}\psi_{i}\left(\theta_i=\frac \pi 3,\phi_i\right)\,,
\;\;\;\;\; i=1,2,3,4.
\label{racc-cond}
\ee
Since we are interested in the orbifold limit in which the spherical caps collapse to zero size,
we consider the conditions (\ref{racc-cond}) as a set of constraints that has to be satisfied
by the wave function $\psi_0$, taking $\psi_i$ as in Eq.~(\ref{spheresolu}), {\it i.e.}
with definite angular momentum. Eq.~(\ref{racc-cond}) becomes indeed, in the orbifold limit, a condition on the
power-behavior of $\psi_R(z)$, $\psi_L(\ov z)$ when $z$ approaches one of the fixed points. Therefore, even if we
would have taken $\psi_i$ as a linear combination of solutions with definite $N_{L,R}^{(i)}$\,, just one of them
would have been responsible of the leading contribution.
Eqs.~(\ref{racc-cond}), in particular, imply that close to each of the 4 fixed points $z_i$,
the holomorphic and anti-holomorphic functions $\psi_{0,R}(z)$ and $\psi_{0,L}(\bar z)$ behave as
\bea
\psi_{0,R}(z) \simeq \a\a (z-z_i)^{\lambda_R^{(i)}} \,, \nn \\
\psi_{0,L}(\bar z) \simeq \a\a (\bar z-\bar z_i)^{\lambda_L^{(i)}}\,,
\label{zi-lambda}
\eea
where
\bea
\lambda_R^{(i)} = \a\a 2 N_R^{(i)} + \frac{\kappa_i+1}2\,, \;\;\;\;\;\; {\rm with} \;\; N_R^{(i)} \geq 0 \,, \nn \\
\lambda_L^{(i)} = \a\a - 2 N_L^{(i)} - \frac{\kappa_i-1}2\,, \;\;\; {\rm with} \;\; N_L^{(i)} \leq 0 \,.
\label{lambda-i}
\eea
Since the $\kappa_i$ are always odd integer numbers, as it is clear from their definition given in
Eq.~(\ref{ki}), $\lambda_{L,R}^{(i)}$
are always integer numbers, implying that $\psi_{0,R}$ and $\psi_{0,L}$ have poles or zeros of order
$|\lambda_{L,R}^{(i)}|$ at $z=z_i$.

An holomorphic function that is periodic for $z\rightarrow z+1$ and $z\rightarrow z+\tau$
is called an elliptic function and fulfills many interesting properties (see {\it e.g.} \cite{WW}).
All the properties of elliptic functions we will consider are actually easily generalized
for non-periodic functions as well (namely, to the cases in which
$(t_1,t_2)\neq (0,0)$ in Eq.~(\ref{per-cond})); hence,
in the following we will consider the general case with arbitrary $t_1$ and $t_2$ and,
with an abuse of language, continue to refer to elliptic functions.
Our strategy in finding the zero modes $\psi_{0,R}(z)$ and $\psi_{0,L}(\bar z)$ will then be the following:
we look for elliptic functions on the covering torus $T^2$ of the original orbifold whose
behavior, close to the fixed points $z_i$, is as in Eq.~(\ref{zi-lambda}).
Once found, we simply identify $\psi_{0,R}(z)$ and $\psi_{0,L}(\bar z)$ as the restriction to the
flat region of ${\cal M}$ of these functions.

The first important property of elliptic functions regards the number of their (simple) poles and zeroes,
counted including their degree of multiplicity (namely a pole/zero of order $n$ is equivalent to
$n$ simple poles/zeroes). The difference between the number of poles and zeroes of an elliptic
function $f(z)$ inside its fundamental domain ${\cal P}$ (a parallelogram in the complex plane whose
vertices are $(z,z+1,z+\tau,z+1+\tau)$, where $z$ is an arbitrary complex number) is always zero.
This is easily established by computing the contour integral $\oint_{\partial{\cal P}}dz f^\prime/f$.
It implies that the coefficients $\lambda_{L,R}^{(i)}$ defined in Eq.(\ref{zi-lambda}) must satisfy
the conditions
\bea
b_R - a_R + \sum_{i=1}^4 \lambda_R^{(i)} = \a\a b_R - a_R + \sum_{i=1}^4  2 N_R^{(i)} +2 + 2\kappa = 0 \,, \nn \\
b_L - a_L +\sum_{i=1}^4  \lambda_L^{(i)} = \a\a b_L - a_L - \sum_{i=1}^4 2 N_L^{(i)} +2 - 2\kappa = 0\,,
\label{condiI}
\eea
where $b_{R,L}$ and $a_{R,L}$ represent the number of possible additional zeroes
and poles that $\psi_{0,R}$ and $\psi_{0,L}$ might have on ${\cal P}$.
Assume, for the moment, that the points $z_i$ are the only poles or zeroes
of $\psi_{0,L/R}$ so that $b_R=a_R=b_L=a_L=0$. We will come back to discuss the validity
of this assumption later. In this case,
since $N_R^{(i)} \geq 0$ and $N_L^{(i)} \leq 0 $, we can immediately establish the important
result that for $\kappa>0$ ($\kappa<0$) no right-handed (left-handed) zero modes exist.
Using the Atiyah-Singer index theorem (\ref{AS-index}), this implies ($n_L=|\kappa|,\; n_R=0$)
or $(n_L=0,\; n_R=|\kappa|$). For $\kappa=0$, the conditions (\ref{condiI}) do not admit any solution
and thus one has $n_L=n_R=0$.
For given $\kappa_i$, the wave functions $\psi_{0,R}$ and $\psi_{0,L}$ are given by
the elliptic functions (holomorphic and anti-holomorphic) with the behavior as in
Eq.~(\ref{zi-lambda}), with $\lambda^{(i)}_{L,R}$ subject only to the constraint (\ref{condiI}).

Another important property of elliptic functions regards the location of their
poles and zeroes. Denote by $a_i$ and $b_i$ ($i=1,\ldots,n$) the
location in the complex plane of the poles and zeroes of an elliptic function,
where a pole or zero is repeated a number of times equal to its degree of multiplicity.
By computing the contour integral $\oint_{\partial{\cal P}}dz z f^\prime/f$, one can easily
establish that
\be
\sum_{i=1}^n \Big(a_i - b_i\Big) = \Big(p+\frac{t_2}2\Big) + \tau \Big(q+\frac{t_1}2\Big)\,,
\label{ai-bi}
\ee
where $p,q$ are arbitrary integers. By appropriately choosing ${\cal P}$, one can always
set $p=q=0$.

Modulo a constant, the form of an elliptic function is uniquely determined once
the location of its poles and zeroes is known.
In terms of theta functions (see the Appendix B), and taking $p=q=0$ in Eq.~(\ref{ai-bi}), one has
\be
f(z) = {\cal N} \, e^{i\pi z t_1} \prod_{i=1}^n \frac{\theta_1(z-b_i|\tau)}{\theta_1(z-a_i|\tau)}\,,
\label{fz-exp}
\ee
with ${\cal N}$ a normalization factor. Eq.~(\ref{fz-exp})
can be verified using Eq.~(\ref{ai-bi}) and the periodicity properties of $\theta_1(z)$,
given in Eq.~(\ref{theta1-per}).
This is the generic form of a zero mode $\psi_{0,R}$, where the poles $a_i$ and zeroes
$b_i$ are fixed according to Eq.~(\ref{zi-lambda}). Similarly, the zero modes $\psi_{0,L}$
are given by the complex conjugate of Eq.~(\ref{fz-exp}).
A careful reader will quickly recognize that for a given set of integers $\kappa_i$, the number
of allowed sets for the $\lambda^{(i)}_{L,R}$'s is typically greater than $\kappa$, in apparent
contradiction with the index theorem and our assumptions. This conflict is resolved
by noticing that generally the functions (\ref{fz-exp}) are not all independent from each other.
In fact, although we have not been able to find a general proof, for values of $|\kappa|$ up
to $4$ we have explicitly checked, using repeatedly Eqs.~(\ref{theta-relations}), that the
number of independent solutions is always $|\kappa|$, as it should. We expect that this is true
for all values of $|\kappa|$.

After this lengthy, but necessary, mathematical digression, we are ready to analyze in detail
the nature of the fermion zero modes on ${\cal M}$, for given values of $\kappa_i$.
First of all, we notice that in the limit of vanishing spherical caps,
the wave functions $\psi_{0,R(L)}$ will be localized at the fixed point $z=z_i$ corresponding
to the pole of maximum degree.\footnote{There could be more than one pole
with the maximum degree. In this case, the wave functions are localized on more than
one fixed point, as shown in the following.}
Indeed, the normalization factor ${\cal N}$ in Eq.~(\ref{fz-exp}) is chosen
such that
\be
\int_{{\cal M}} |f(z)|^2 d^2z = 1\,.
\label{normalization}
\ee
If the integral diverges due to the presence of various poles at $z=z_i$, the factor
${\cal N}\rightarrow 0$ so that the normalization (\ref{normalization}) is satisfied.
Consequently, the wave function vanishes everywhere but in a small region around the pole of maximum
degree. These states correspond then to localized fermions in the orbifold limit.
A bulk zero mode, on the contrary, is given by the constant wave function, namely when
$\lambda^{(i)}_{L,R}=0$, $\forall i$.
It is simple to see from Eqs.~(\ref{lambda-i}) that, as expected from Eqs.~(\ref{or-cond}) and (\ref{per-cond}),
this condition can be realized only when $t_1=t_2=0$ and, depending on whether $p=1$ or $p=-1$
a solution is possible only for the left-handed or right-handed fields, respectively.
However, since $N_R^{(i)}\geq 0$ and $N_L^{(i)}\leq 0$, the condition $(t_1,\,t_2)=(0,\,0)$ in Eq.~(\ref{pi}) is not
sufficient for the zero mode to exist, one must also assume the integers $n_i$ in Eq.~(\ref{ki})
to have the same sign as $p$.
This is not surprising; it is analogous to what we found in Sect.2 when studying the case of the
non-compact $\CC/\Z_N$ orbifold. Note that the absence of a bulk zero mode when some $n_i$ has the ``wrong'' sign
is not the signal of an inconsistency in the resolution.
We interpret it as a possible resolution of an orbifold
in which bulk fermions are not present, and we have localized fields only.

If $(t_1,\,t_2)\neq(0,\,0)$, no bulk zero modes are present, independently on the signs of $n_i$.
It is then not possible, by an analysis of zero modes only, to establish which are the right signs of the
$n_i$'s for the bulk fermion to be present in the corresponding orbifold model.
Our conjecture is that, in order for bulk fields to exist, the sign of each
$n_i$ must be the same as the one of the corresponding $p_i$, {\it i.e.} $\kappa_i=p_i|\kappa_i|$.
In the following, we will focus on this case
and a discussion on the validity of our conjecture is postponed to the end of Sect.4, where we will present an argument
showing that massive states whose mass does not diverge in the orbifold limit can only be present on $\mc{M}$
if each $n_i$ has the same sign of the corresponding $p_i$.

We now consider in detail some examples that will help to clarify
the considerations previously done. Take $\kappa=-2$, with $p=-1$, $t_{1,2}=0$, $n_1=-1$,
$n_2=n_3=n_4=0$ in Eqs.(\ref{ki}, \ref{pi}).  Since $\kappa<0$, we know from Eqs.~(\ref{condiI}) and
(\ref{AS-index}) that there should be $2$ right-handed and no left-handed fermions.
Plugging in Eq.~(\ref{lambda-i}) the values of $\kappa_i$ obtained from Eq.~(\ref{ki}),
one has $\lambda_R^{(1)}= 2N_R^{(1)}-2$, $\lambda_R^{(j)}=2N_R^{(j)}$
($j=2,3,4$). There are four possible configurations that solve the constraint (\ref{condiI}),
associated to the four cases in which one $N_R$ equals $1$ and the others are vanishing.
When $N_R^{(1)}=1$, all $\lambda_R^{(i)}$ vanish and we have the constant zero mode.
If $N_R^{(1)}=0$ and anyone of the remaining $N_R$ is equal to $1$: $N_R^{(j)}=1$ ($j=2,3$ or $4$),
the corresponding solutions are localized around $z=0$
(that is a second order pole) and in the limit of vanishing
spherical caps read as
\be
\psi_{0j,R}(z) = {\cal N}_j\, \Bigg[\frac{\theta_j(z|\tau)}{\theta_1(z|\tau)}\Bigg]^2 \,.
\label{psi-expl}
\ee
Eqs.(\ref{psi-expl}) are obtained by the general formula (\ref{fz-exp}) using the various
relations between the theta functions reported in Appendix B. There are $4$ different
zero modes, namely the constant and the three $\psi_{0j,R}$ of Eq.(\ref{psi-expl}), but it is not
difficult to verify, using the identities (\ref{theta-relations}), that only two solutions
are independent, as expected for $\kappa=-2$. In the orbifold limit, the $3$ functions
$\psi_{0j,R}$ are all indistinguishable from each other.
The two physically distinct solutions can thus be taken to be
the constant and the symmetric combination $\psi_{02,R}+\psi_{03,R}+\psi_{04,R}$.
Due to the double pole at the origin, the latter wave function, in the
orbifold limit, gives rise to a Dirac-like distribution, localized
at $z=0$.
Summarizing, we have found that for
$\kappa=-2$, $p=-1$, $t_{1,2}=0$, $n_1=-1$, $n_2=n_3=n_4=0$, there are two massless right-handed
fermions, one with a wave function localized at $z=0$, and the other
with a wave function constant on ${\cal M}$.
In the orbifold limit, this configuration is nothing else that the one obtained by having
a bulk fermion with parity as in Eq.(\ref{or-cond}) and one right-handed fermion with the same
$U(1)$ charge, localized at $z=0$.
Higher values of $|\kappa|$ can similarly be studied. The reader can check that for
$\kappa=-3$, $p=-1$, $t_{1,2}=0$, $n_1=n_2=-1$, $n_3=n_4=0$, one has 3 independent solutions.
These can be taken to be the constant (a bulk mode),
$[\theta_2^2(z)+\theta_3^2(z)+\theta_4^2(z)]/\theta_1^2(z)$ (a mode localized
at $z=0$) and $[\theta_1^2(z)+\theta_3^2(z)+\theta_4^2(z)]/\theta_2^2(z)$ (a mode localized at $z=1/2$).
Here and in the following, for simplicity, we omit the dependence on $\tau$ of the theta
functions $\theta_i$.
For $\kappa=-3$, $p=-1$, $t_{1,2}=0$, $n_1=-2$, $n_2=n_3=n_4=0$, the $3$ independent solutions
are the constant (bulk mode), $[\theta_2^2(z)+\theta_3^2(z)+\theta_4^2(z)]/\theta_1^2(z)$
(first mode localized at $z=0$) and
$\{[\theta_2^2(z)+\theta_3^2(z)+\theta_4^2(z)]/\theta_1^2(z)\}^2$ (second mode localized at $z=0$).
The general case should now be clear. The wave functions are only labeled by the degree and the location of the
 maximum pole, in the sense that among all the functions with a certain maximum pole,
only one is linearly independent, the others being linear combinations of it with other functions with maximum pole of
lower degree.
By assuming this, one is able to demonstrate that in general, as confirmed by many examples,
for $|\kappa|=1+\sum_{i=1}^4 |n_i|$, one has one bulk mode and $|n_i|$ fermions localized at $z=z_i$.
They lead to four dimensional fermions all of the same chirality, depending on the sign of $\kappa$ and of the $n_i$'s.

For $(t_1,t_2)\neq (0,0)$, as we said, no constant wave function can appear and all zero modes
give rise to localized chiral fermions.
For instance, take $p=-1$, $t_1=1$, $t_2=0$. The corresponding $p_i$, as defined in Eq.~(\ref{pi}), are given by
$p_1=-1$, $p_2=+1$, $p_3=+1$, $p_4=-1$. According to our rule for which $\kappa_i=p_i|\kappa_i|$,
we take $n_{1,4}=-|n_{1,4}|$ and $n_{2,3}=+|n_{2,3}|$. Consider, for instance, the case
$|\kappa|=-\kappa=|n_1|+|n_4|-|n_2|-|n_3|=1$, realized with $|n_1|=1$,
$|n_{2,3,4}|=0$. From Eq.~(\ref{lambda-i}), we see that the only (right-handed)
wave function (note that $N_{R}^i=0$ $\forall i$, if $\kappa=-1$)
has one pole of order $2$ at $z_1$ and corresponds to a state localized at $z=0$. If we consider again $\kappa=-1$,
realized now with $|n_{1,4}|=1$, $|n_{2}|=0$ and $|n_3|=1$ (or $|n_{3}|=0\,, |n_2|=1$), a new feature appears.
{}From Eq.~(\ref{lambda-i}) we see that the wave function has now two poles of order $2$ at $z_1$ and $z_{4}$.
This case, which is never realized when $t_{1,2}=0$,
corresponds to a wave function which is localized at \emph{two} different points. The reader can check that,
for any configuration
with $\kappa=-1$, the field is localized at $z_1$ if $|n_1|>|n_4|$, at $z_2$ in the opposite case and on
both when $|n_1|=|n_4|$.
The case $\kappa=+1$ is similar:
we have one left-handed field localized at $z_2$ or $z_3$, depending on whether $|n_2|$ is greater or
smaller then $|n_3|$. When they are equal, as before, we have a double localization.
The case $p=+1$ is analogous, as well as the various cases where $(t_1\,,t_2)=(0\,,1)$ or $(1\,,1)$.
The general results for any $\kappa$ are shown in Table 1.

\begin{table}[t]
\begin{center}
\begin{tabular}{|c|c|c|c|c|}
\hline ($t_1,t_2$)& $z_1$ & $z_2$ & $z_3$ & $z_4$  \\
\hline
& $|n_1|$ & $|n_2|$ & $|n_3|$ & $|n_4|$  \\ ($0,0$) & & & &  \\
& --  &  -- &  --  & -- \\ \hline
& $\frac{|\kappa|+|n_1|-|n_4|}2$ & $0$ & $0$ & $\frac{|\kappa|+|n_4|-|n_1|}2$  \\ ($1,0$) & & & &  \\
& $0$ &  $\frac{|\kappa|+|n_2|-|n_3|}2$ & $\frac{|\kappa|+|n_3|-|n_2|}2$ & $0$ \\ \hline
& $\frac{|\kappa|+|n_1|-|n_2|}2$ & $\frac{|\kappa|+|n_2|-|n_1|}2$ & $0$ & $0$  \\ ($0,1$) & & & &  \\
& $0$ & $0$ & $\frac{|\kappa|+|n_3|-|n_4|}2$ & $\frac{|\kappa|+|n_4|-|n_3|}2$ \\ \hline
& $\frac{|\kappa|+|n_1|-|n_3|}2$ & $0$ & $\frac{|\kappa|+|n_3|-|n_1|}2$ & $0$  \\ ($1,1$) & & & &  \\
& $0$ & $\frac{|\kappa|+|n_2|-|n_4|}2$ & $0$ & $\frac{|\kappa|+|n_4|-|n_2|}2$ \\ \hline
\end{tabular}
\caption{\footnotesize{The number of fermions localized at each of the fixed points $z_i$, for the different values
of $(t_1,\,t_2)$, when the sign of each $n_i$ is the same as the corresponding $p_i$.
In each column, the upper value corresponds to the case in which $p$ has the same sign as
$\kappa$, the lower one to the case of opposite sign. Of course, the second case cannot be realized for
$(t_1,\,t_2)=(0,\,0)$.
When a number in the table is negative, it has to be replaced with $0$, while the other non-vanishing number on the same raw
must be replaced with $|\kappa|$.
Moreover, when it is positive but semi-integer, its integer part gives the number of states localized at
the corresponding fixed point,
while the extra $1/2$ represents a state which is localized at two points. It is understood that
for $\kappa>0$ and $\kappa<0$, the table refers to left-handed and right-handed fields,
respectively.}}
\end{center}
\end{table}

Before concluding this section, there is still a point to be discussed, regarding our previous assumption
that the points $z_i$ are the only
points where the wave functions $\psi_{0,R/L}$ have zeroes or poles, namely that
$b_R=a_R=b_L=a_L=0$ in Eqs.~(\ref{condiI}).
A posteriori, the fact that we have always found all the $|\kappa|$ independent zero mode
solutions to the Dirac equation, in agreement with the index
theorem, provides a strong consistency check on the validity of the above assumption.
However, one cannot exclude that the possible presence of additional poles and
zeroes in the flat region of ${\cal M}$ might led to extra left and right-handed zero modes,
equal in number, so that the index theorem would still be respected.
This possibility is excluded by noting that no additional poles can be present on ${\cal M}$,
otherwise necessarily there would be zero modes that get localized at these points.
In the orbifold limit, this would not make sense since the only singular points are the points $z_i$
and thus we are led to conclude that $a_R=a_L=0$.
We now see from Eqs.~(\ref{condiI}) that, for given $b_L,b_R$,
right-handed (left-handed) zero modes can exist for $\kappa\leq -1-b_R/2$
($\kappa\geq 1+b_L/2$).
For any choice of $b_L$, $b_R$ and $\kappa$, we never get left and right modes at the same time,
as required by the index theorem. Hence, no zero modes escaped to our analysis.
Moreover, since for $\kappa=\pm 1$ we must have one zero mode, we conclude that $b_R=b_L=0$,
proving the validity of our initial assumption.

\section{$S^1/\Z_2$ as the degenerate limit of $T^2/\Z_2$}

Consider a $T^2/\Z_2$ orbifold with complex structure
$\tau=it$ ($t$ real) and fermions periodic for $z\rightarrow z+\tau$ ($t_2=0$ in Eq.~(\ref{T-6D})).
In the degenerate limit $t\rightarrow0$, this set-up degenerates
to a one dimensional $S^1/\Z_2$ orbifold with periodic (if $t_1=0$)
or anti-periodic (if $t_1=1$) fermions on the covering circle $S^1$.
In this limit, the two-dimensional orbifold (see Fig.~4) becomes
a one-dimensional segment of length $1/2$, where the $z_1$ and $z_4$,
as well as the $z_2$ and $z_3$, fixed points collapse to a single point.
The bulk of the $S^1/\Z_2$ orbifold, which corresponds to the flat region of $\mc{M}$
 when the degeneration limit is taken, is well reproduced by a cylinder of
length $L=1/2$ and radius $r\rightarrow0$, which is indeed a flat two dimensional space degenerating to a
segment of length $1/2$.
The two fixed points which we call ``$(1)$'' and ``$(2)$'', on the contrary,
correspond to the non-flat regions of $\mc{M}$,  the $(1)\cup\,(4)$ and
 $(2)\cup\,(3)$ sets, respectively.
We represent them as two half-spheres with a suitable gauge
connection on it, such that the total
(gauge$+$Lorentz) holonomies of their boundaries are trivial (look at Eq.~(\ref{Pi}) for $t_2=0$),
being the product of those around $z_1$ and $z_4$ and around $z_2$ and $z_3$, respectively.

To summarize, the ${\cal C}^1$ resolution of the $T^2/\Z_2$ orbifold has suggested us that a cigar-like surface
(which we denote as $\mc{C}$), shown in Fig.~5, could reproduce the $S^1/\Z_2$ orbifold if we put on the two
half-spheres a non-trivial gauge connection which makes trivial the gauge$+$Lorentz holonomy of a circuit around
the cylinder. Even if the $S^1/\Z_2$ orbifold is simply a one-dimensional space with boundaries, and therefore does not
need of any resolution,
it is in any case interesting to see whether localized chiral fermions naturally arise when it is seen as the limit of the
two-dimensional cigar and to classify the localization pattern we can obtain. This is the subject of the next subsection.
\begin{figure}[t]
\begin{center}
\begin{picture}(200,160)
\psfrag{g1}{\footnotesize $\gamma_1$}
\psfrag{g2}{\footnotesize $\gamma_2$}
\put(50,0){\includegraphics*[width=3cm]{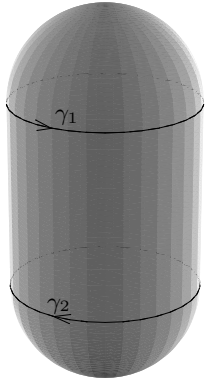}}
\end{picture}
\caption{\footnotesize{The cigar-like surface $\mc{C}$, which reproduces the $S^1/\Z_2$
 orbifold when the radius $r$ of the two half-spheres goes to zero.
The oriented curves $\gamma_{1,2}$, on which the two
half-spheres intersect the cylinder, are also shown.}}
\end{center}
\protect\label{fig.5}
\end{figure}

\subsection{Chiral fermions on the cigar}

The cigar (see Fig.~5) is composed by $3$ sets; the two half-spheres (``$(1)$'' and ``$(2)$'')
and the flat cylindrical region connecting them (which we call the ``$(0)$'' set).
We introduce spherical coordinates $(\theta_i,\phi_i)$ on the two half-spheres, such
that $\theta_i\in[0,\pi/2]$, $\phi_i\in[0,2\pi)$. On the cylinder $(0)$, we introduce cylindrical
coordinates $(z,\alpha)$ with $\alpha\in[0,2\pi)$ and $z\in[-L/2,L/2]$, $L$ being the length of the
``resolved'' $S^1/\Z_2$ orbifold.
The set $(0)$ intersects with the set $(1)$ on the oriented
circle $\gamma_1$ (corresponding to $z=-L/2$ in cylindrical coordinates) and with ``$(2)$'' on
$\gamma_2$ (on which it is $z=+L/2$). The two circles $\gamma_{1,2}$ are the equators $\theta_i=\pi/2 $ of
the two spheres, where one has, respectively, $\phi_1=\alpha$ and
$\phi_2=-\alpha$.

On the $(0)$ set, both the gauge and spin connections vanish, whereas
on the two half-spheres the spin connection $\omega_i$ is, as usual, given by Eq.~(\ref{omega-i}) and
a gauge field $A_i=\tilde\kappa_i/2\omega_i$ is present. The holonomies around the circuits $\gamma_i$
are given by
\be
W_i=\ds{e^{-i\oint_{\gamma_i}(A_i+\frac12\sigma_3\omega_i)}=
e^{-\frac{i}2(\tilde \kappa_i+\sigma_3)\oint_{\gamma_i}\omega_i}=
e^{-\frac{i}2(\tilde \kappa_i+\sigma_3)\int_i R}=e^{-\pi i(\tilde \kappa_i+\sigma_3)}}\,.
\label{holo-5d}
\ee
As discussed above, in order for the cigar to reproduce the $S^1/\Z_2$ orbifold, both holonomies
have to be trivial ($W_1={W_2}^{-1}=1$) and hence both $\tilde\kappa_i$'s must be odd integers.

In the following, we will be interested in comparing the localization pattern obtained from the cigar with
the one obtained by $\mc{M}$ for $T^2/\Z_2$, when the degeneration limit is taken. For this comparison to be done,
the integrals of the gauge field-strength on the $(1)$ and $(2)$ sets must be given by the sum of the
integrals on the $(1)$ and $(4)$ and on the $(2)$ and $(3)$ sets of $\mc{M}$, respectively.
Then, we must have
\bea
\tilde \kappa_1 = \a\a \frac{\kappa_1+\kappa_4}2=p+2(n_1+n_4)\,, \nn \\
\tilde \kappa_2 = \a\a \frac{\kappa_2+\kappa_3}2=(-)^{t_1}p+2(n_2+n_3) \,,
\label{k-ktilda}
\eea
having made use of Eq.~(\ref{ki}).

Consider now the massless Dirac equation on the cigar.
On the two half-spheres, the solutions $\psi_{i,R}$ and $\psi_{i,L}$
are the by-now well-known Eqs.~(\ref{spheresolu}), with
$N_R^{(i)}\ge0\,, N_L^{(i)}\le0$ ($i=1,2$).
On the cylinder, the Dirac equation for the fermion field $\psi_0$ reads
\be
\left[i\sigma_1\partial_z+i\sigma_2\frac1{r}\partial_{\alpha} \right]\psi_{0}=0\,,
\ee
where $r$ is the radius of the half-spheres.
Modulo a constant, the solutions are arbitrary functions of $z/r \pm i \alpha$ for
right and left-handed fermions. Precisely like in the case of the cone $\CC/\Z_N$,
the ones that match Eqs.~(\ref{spheresolu}) are simple monomials,
\bea
\psi_{0,L}=e^{\rho_L(-\frac{z}{r}+i\alpha)}\,,\nn \\
\psi_{0,R}=e^{\rho_R(+\frac{z}{r}+i\alpha)}\,.
\eea
To connect these solutions to those on the half-spheres, we use the fact that
\be
\begin{array}{c} \omega_{0}|_{\gamma_i}= \omega_{i}(\theta_i=\frac \pi2)-d\phi_{i}\,, \\
A_{0}|_{\gamma_i}= A_{i}(\theta_i=\frac \pi2)-\frac{\tilde \kappa_i}2 d\phi_{i}\,,
\end{array}
\label{cigarI}
\ee
since $\omega_0=A_0=0$. Eqs.~(\ref{cigarI}) lead to the following identifications
for the wave functions:
\be
\begin{array}{c}
\psi_{0}(z=-\frac{L}2,\alpha=\phi_1)=\ds{e^{\frac{i}2 \phi_{1}(\tilde \kappa_i+\sigma_3)}}
\psi_{1}(\theta_1=\frac \pi 2, \phi_1)\,, \\ \vspace{-0.4cm} \\
\psi_{0}(z=\frac{L}2,\alpha=-\phi_2)=\ds{e^{\frac{i}2 \phi_{2}(\tilde \kappa_i+\sigma_3)}}
\psi_{2}(\theta_2=\frac \pi 2, \phi_2)\,.
\end{array}
\label{hyp2}
\ee
In this way, $\rho_{L,R}$ are expressed in terms of the fluxes $\tilde \kappa_{1,2}$ and the angular momenta
$N_{L,R}^{1,2}$ of the solutions on the half-spheres:
\bea
\rho_L=\frac{\tilde\kappa_1-1}2+N^{(1)}_L=-\bigg(\frac{\tilde\kappa_2-1}2+N^{(2)}_L\bigg)\,,\nn \\
\rho_R=\frac{\tilde\kappa_1+1}2+N^{(1)}_R=-\bigg(\frac{\tilde\kappa_2+1}2+N^{(2)}_R\bigg)\,.
\label{l-cigar}
\eea
By using Eqs.~(\ref{l-cigar}) and (\ref{ang-cond}), one finds $0\leq N_R^{(1)} \leq -\tilde\kappa -1$ and
$1-\tilde\kappa \leq N_L^{(1)} \leq 0$, where $\tilde\kappa=(\tilde \kappa_1+\tilde \kappa_2)/2$
is the total flux on ${\cal C}$.
Once again, for $\tilde\kappa>0$ only left-handed states are present ($n_L=\tilde\kappa$, $n_R=0$) and only right-handed
ones for $\tilde\kappa<0$ ($n_L=0$, $n_R=|\tilde\kappa|$).
According to the value of $\rho_{L,R}$, in the limit $r\rightarrow 0$, the zero modes
are localized at $z=L/2$ ($\rho_{R}>0,\rho_{L}<0,$), at $z=-L/2$ ($\rho_{R}<0,\rho_{L}>0$),
or they are bulk modes ($\rho_{L,R}=0$). If $\kappa_1$ and $\kappa_2$ have the same signs,
we thus get a bulk mode and fields localized at $z=\pm L/2$, whose number depends on the
values of $\tilde\kappa_1$ and $\tilde\kappa_2$; if they have opposite signs, no massless bulk mode is present,
and all the states are localized at $z=L/2$ or $z=-L/2$. They have all the same chirality,
depending on the sign of the resulting $\tilde \kappa$.

As a result, there is a complete agreement between the localization pattern which
we derived in this section with that shown in the first two lines of Table~1.
Indeed, being $\tilde\kappa_{1,2}$ related to the values of $\kappa_{1,2,3,4}$ as in Eq.~(\ref{k-ktilda})
and the signs of $\kappa_{1,2,3,4}$ assumed to be the same of the corresponding $p_i$'s defined in
Eq.~(\ref{pi}), the case in which $\tilde\kappa_{1}$ and $\tilde\kappa_{2}$
have the same sign corresponds to the first line of the table, and the case in which
they have opposite sign to the second line. As it can be seen, there
is agreement between the number of states localized at $(1)$ and the sum of those on $z_1$ and $z_4$, while
those on $(2)$ are the ones on $z_2$ plus those on $z_3$.

It is not clear, however, that the cigar $\mc{C}$ is completely equivalent to $\mc{M}$ in the degeneration
limit. The correspondence seems to fail in the case in which some $\kappa_i$ in $\mc{M}$ have not the sign of the
corresponding $p_i$'s, {\it i.e.} in the case in which no bulk fields have been introduced in the
orbifold theory. An example of the above fact is
that one can have a bulk zero mode on ${\cal C}$, despite
on ${\cal M}$ no bulk mode is present. This is technically explained by noting that when ${\cal M}$ collapses to
${\cal C}$, a given combination of localized zero modes on ${\cal M}$
can give rise to a constant mode. This can be verified by studying the limit
$\tau\rightarrow 0$ of the wave functions (\ref{fz-exp}), but we will not enter here in a detailed analysis.
It can suffice to say that when $t_1=1$, $t_2=0$, one can have a pole at a fixed point and a zero
at another fixed point that, in the limit in which the two fixed points collapse to a single point,
compensate each other resulting in a constant wave function.
This is what happens, for instance, for the wave functions $\theta_2(z)/\theta_3(z)$ and
$\theta_3(z)/\theta_2(z)$ that arise by taking $\kappa=-2$ with $p=-1$ and $t_1=1$ in Eq.~(\ref{pi}) and
$n_1=n_4=0$ and $n_2=n_3=-1$ in Eq.~(\ref{ki}).
On ${\cal C}$, this configuration
has $\tilde \kappa_1=-1,\tilde \kappa_2=-3$ and gives rise to one bulk constant mode and one mode localized at $z=L/2$.
It would be interesting to better formulate the correspondence between the cigar $\mc{C}$
and the space  $\mc{M}$ to better understand this class of configurations.

Independently on the correspondence with $T^2/\Z_2$, all the results we obtained from the degeneration limit
of the cigar $\mc{C}$ can be interpreted in terms of a 5D fermion $\chi$ on the orbifold $S^1/\Z_2$.
The most general boundary conditions one can have are
\bea
\chi(-z-\pi R) = \a\a \eta \gamma_5 \chi(z)\,, \nn \\
\chi(-z+\pi R) = \a\a \eta' \gamma_5 \chi(z)\,.
\label{pro-5d}
\eea
In Eq.(\ref{pro-5d}), $\eta,\eta^\prime=\pm 1$ represent the two local projections around the two fixed points
$z_1=-\pi R/2$ and $z_2=\pi R/2$, where $R=L/\pi$ is the radius of the covering circle of $S^1/\Z_2$.
The periodicity of the fermion around the covering circle $S^1$ is determined for consistency,
once $\eta$ and $\eta^\prime$ are fixed.
The fermion $\chi$ is periodic if $\eta=\eta^\prime$, antiperiodic if $\eta=-\eta^\prime$.
In the latter case, the space is effectively an $S^1/(\Z_2\times\Z_{2}')$ orbifold.
If $\eta=\eta^\prime=\pm 1$, a left or right-handed massless fermion is present; on the contrary,
for $\eta=-\eta^\prime=\pm 1$, all states are massive.

These configurations of bulk fermions are reproduced by taking $|\tilde\kappa_1|=\eta\tilde\kappa_1$ and
$|\tilde\kappa_2|=\eta'\tilde\kappa_2$.
In next subsection we will support the above correspondence by a study of the
massive fermions on ${\cal C}$.

\subsection{Light massive fermions on the cigar}

Depending on the relative sign of the gauge fluxes $\tilde\kappa_{1,2}$ on the two spherical caps,
we have seen that
the cigar ${\cal C}$ should mimic, in the limit in which its thickness goes to zero, a segment
with $S^1/\Z_2$ or $S^1/(\Z_2\times\Z_{2}')$ boundary conditions.
As a further consistency check of this statement, we consider here
the massive Dirac equation on this space, showing that the mass-spectrum and the wave functions are
indeed what is expected for one dimensional orbifolds.

The massive Dirac equation on the ($0$) set parameterizing the cylinder is trivial. Taking the ansatz
\be
\psi_{0,L(R)}=f_{L(R)}^0(z)\ds{e^{in\alpha}}\,,
\label{par0}
\ee
with $n$ integer, the generic solution is
\be
\left\{
\begin{array}{l}
f_{R}^0=\mathcal{N}^+\ds{e^{ip_n z}}+\mathcal{N}^-\ds{e^{-ip_n z}}
\\
f_{L}^{0} =\frac{1}{m}\Bigg[\bigg(p_n-\frac{in}r\bigg)  \mathcal{N}^-{\ds e^{-ip_n z}}-
\bigg(p_n+\frac{in}r\bigg) \mathcal{N}^+{\ds e^{+ip_n z}}\Bigg]
\end{array}
\right.\,,
\label{cy-sol}
\ee
where $p_n$ is defined by
\be
p_n=\sqrt{m^2-\left(\frac{n}{r}\right)^2}\,.
\label{m-cyl}
\ee
In order to perform our consistency check, it is not needed to compute the full spectrum on
the cigar for $r$ finite, but just the masses and wave functions of the states with finite mass
in the $r\rightarrow 0$ limit, namely with $n=0$ in Eq.~(\ref{m-cyl}).
In the following, we will thus focus on these states only.
According to Eqs.~(\ref{hyp2}) and (\ref{par0}), the $\phi_i$-dependence of the
solutions $\psi_{i,L(R)}$ on the two spherical caps are fixed. Parametrizing them as
\be
\psi_{i,L}=f_{L}^i(\theta_i)\ds{e^{iN_L^{(i)}\phi_i}}\,,
\;\;\;\;\;\psi_{i,R}=f_{R}^i(\theta_i)\ds{e^{iN_R^{(i)}\phi_i}}\,,
\label{pm}
\ee
we have
\bea
N_R^{(i)} = \a\a -\frac{\tilde \kappa_i+1}2\,, \nn \\
N_L^{(i)} = \a\a -\frac{\tilde \kappa_i-1}2\,.
\label{N-f}
\eea
The values of $f_{L,R}^i$ at $\theta_i=\pi/2$ are also fixed:
\bea
\left\{
\begin{array}{l}
f_{R}^1(\theta_1=\pi/2)=\mathcal{N}^+\ds{e^{-i\frac{mL}2}}+\mathcal{N}^-\ds{e^{i\frac{mL}2}}
\\
f_{L}^1(\theta_1=\pi/2)=\mathcal{N}^-\ds{e^{i\frac{mL}2}}-\mathcal{N}^+\ds{e^{-i\frac{mL}2}}
\end{array}
\right.\,,\nn \\
\left\{
\begin{array}{l}
f_{R}^2(\theta_2=\pi/2)=\mathcal{N}^+\ds{e^{i\frac{mL}2}}+\mathcal{N}^-\ds{e^{-i\frac{mL}2}}
\\
f_{L}^2(\theta_2=\pi/2)=\mathcal{N}^-\ds{e^{-i\frac{mL}2}}-\mathcal{N}^+\ds{e^{i\frac{mL}2}}
\end{array}
\right.\,.
\label{b.c.}
\eea

The massive Dirac equation on each spherical cap reads
\be
\left\{
\begin{array}{l} i\partial_\theta\psi_{i,L}
+\frac1{\sin{\theta}}\left[\partial_\phi+i\frac{\tilde \kappa_i-1}2\right]\psi_{i,L}
-i\frac{\tilde \kappa_i-1}2\cot{\theta}\psi_{i,L}=mr\ds{e^{i\phi_i}}\psi_{i,R}\\
i\partial_\theta{\psi_{i,R}^i}
-\frac1{\sin{\theta}}\left[\partial_\phi+i\frac{\tilde \kappa_i+1}2\right]\psi_{i,R}
+i\frac{\tilde \kappa_i+1}2\cot{\theta}\psi_{i,R}=mr\ds{e^{-i\phi_i}}\psi_{i,L}
\end{array}
\right.\,,
\label{cal-m}
\ee
where the two phases on the right-hand side of Eqs.~(\ref{cal-m}) arise from the zweibeins
associated with the spin connections (\ref{omega-i}).
Eqs.~(\ref{cal-m}), with the values of $N_{L,R}^{(i)}$ given in Eq.~(\ref{N-f}),
could be mapped to second-order differential equations
of hyper-geometric form, whose solutions are given by hyper-geometric functions.
Luckily, as we will see, the explicit form of these solutions will not be needed.
In the limit $r\rightarrow 0$, for $m$ finite, the r.h.s. of Eq.~(\ref{cal-m})
is negligible and we are effectively back to the usual massless Dirac equation
on the sphere, whose solutions have the form of Eq.~(\ref{pm}) with
$N_R^{(i)}\geq 0$ and $N_L^{(i)}\leq 0$.
For any given choice of $\tilde \kappa_i$'s, only two of the four conditions (\ref{N-f})
can then be fulfilled. For the remaining two, one is enforced to put to zero
the corresponding wave functions in Eq.~(\ref{b.c.}).
In particular, if $\tilde \kappa_i>0$, $f_{R}^i$ must vanish while $f_{L}^i$
can be non trivial; vice versa for $\tilde \kappa_i<0$. On the other hand, if
$\tilde \kappa_{1,2}>0$ and $\tilde \kappa_{2,1}<0$, one has $f_R^{1,2}=f_L^{2,1}=0$.
Eqs.~(\ref{b.c.}) can then be interpreted as the boundary conditions on the wave functions
$f_{R,L}^0(z)$ defined on the segment $z\in[-L/2,L/2]\,$.
It is now immediate to verify that these boundary conditions
are consistent with our orbifold interpretation of the cigar-like surface. For $\tilde \kappa_i>0$, we get
\be
\begin{array}{c}
\mathcal{N}^+=-\mathcal{N}^-e^{imL}=-\mathcal{N}^-e^{-imL}\;\;\;\Rightarrow\;\;\; e^{2imL}=1\;\;\;
\Rightarrow\;\;\; m=\frac{\pi q}{L}\;\;\; q\in\ZZ\,,\\
\left\{
\begin{array}{l}
\psi_{R}^0=2i\mathcal{N}^-e^{-imL/2}\sin{[m(L/2-z)]}\\
\psi_{L}^0=2\mathcal{N}^-e^{imL/2}\cos{[m(L/2-z)]}
\end{array}\right.\,,
\end{array}
\ee
as expected for an $S^1/\Z_2$ orbifold defined from a circle of radius $L/\pi$.
The case $\tilde \kappa_i<0$ is trivially obtained by interchanging $L$ with $R$.

If $\tilde \kappa_1>0$, $\tilde \kappa_2<0$, we get
\be
\begin{array}{c}
\mathcal{N}^+=-\mathcal{N}^-e^{imL}=\mathcal{N}^-e^{-imL}\;\;\;\Rightarrow\;\;\; e^{2imL}=-1\;\;\;
\Rightarrow\;\;\; m=\frac{\pi}{L}\left(q+\frac12 \right)\;\;\; q\in\ZZ\,,\\
\left\{
\begin{array}{l}
\psi_{R}^0=2i\mathcal{N}^-e^{-imL/2}\sin{[m(L/2-z)]}\\
\psi_{L}^0=2\mathcal{N}^-e^{imL/2}\cos{[m(L/2-z)]}
\end{array}\right.\,,
\end{array}
\ee
as expected for an $S^1/(\Z_2\times\Z_{2}')$ orbifold. Again, the case $\tilde k_1<0$, $\tilde k_2>0$
is obtained by interchanging $L$ with $R$.

The results derived in this section provide a strong evidence that, in the degeneration limit $r\rightarrow0$,
the cigar $\mc{C}$ really resembles the $S^1/\Z_2$ orbifold on which one bulk and various localized fields are
present. The study of massive fermions is also useful to support our conjecture that
the condition $|\kappa_i|=p_i\kappa_i$ is necessary and sufficient for a bulk
field on $T^2/\Z_2$ to be present. Indeed, only in this case orbifold massive states
can be reproduced on $\mc{M}$. In the flat part of $\mc{M}$,
the wave function of light states can reasonably be assumed to be  ``slowly varying'' with respect
to the size $\epsilon\rightarrow0$ of the spherical caps.
They must then be ``basically'' constant on the curves $\gamma_i$ (see Fig.~4), implying that the angular
momenta $N_{L,R}^{(i)}$ of the wave functions on the $i$-th spherical cap is fixed, in analogy with Eq.~(\ref{N-f}),
to be
\bea
N_R^{(i)} = \a\a -\frac{\kappa_i+1}4=-\frac{p_i+1}4-n_i\,, \nn \\
N_L^{(i)} = \a\a -\frac{\kappa_i-1}4=-\frac{p_i-1}4-n_i\,,
\label{N-f2}
\eea
having used the definition of $\kappa_i$ in Eq.~(\ref{ki}).
The conditions (\ref{N-f2}) and (\ref{ang-cond}) can be satisfied at the same time only if
the integers $n_i$ have the same sign as the corresponding $p_i$'s.
When this does not happen, we are enforced to put to zero both the left-
and right-handed components of the wave function at the corresponding fixed point where $n_i$ has not
the same sign as $p_i$. Since this never happens for the usual wave functions of massive fermions on $T^2/\Z_2$,
we are led to the conclusion that $\kappa_i=p_i|\kappa_i|$ is a necessary condition
to reproduce bulk fermions on $T^2/\Z_2$.

On the contary, the reader can verify that, when $\kappa_i=p_i|\kappa_i|$, the usual wave functions of the massive states on
the $T^2/\Z_2$ orbifold are consistent with Eq.~(\ref{N-f2}), in the sense that they vanish precisely at those fixed points
for which Eq.~(\ref{N-f2}) cannot be satisfied.

\section{Outlook}

In this paper we have shown how orbifold field theories, in particular
bulk and localized fermion fields, precisely arise as limits of fermions
on resolved spaces. We think that this represents an important point in support
of the recent bottom-up approach to extra dimensions, in which a microscopic
fundamental theory is generally unknown. The same analysis has however
pointed out that not all possible fermion field configurations admit
a resolution in the sense just explained. On the contrary, we have found that
only a limited class can be defined on the resolved space.
We think that this can pose a strong constraint that should be taken into account
in studying models in extra dimensions,\footnote{In general, strictly speaking, our analysis
does not exclude that from more complicated resolutions one can get
different massless fermion configurations. The full agreement
of our ``cut and paste'' procedure with the ${\cal C}^\infty$ resolution of subsection
2.3, however, seems to suggest that our results are universal.} particularly
for models on $T^2/\Z_2$ orbifolds
and their degenerate limits as $S^1/\Z_2$ orbifolds.\footnote{Notice that there exist
models on $S^1/\Z_2$ that cannot be seen as the degeneration limit of a $T^2/\Z_2$ theory.
In this case, our results might not apply.}

The results of this paper are preliminary in many respects and several important
generalizations should be considered in the next future.
{}First of all, non-abelian gauge fields should be added, so that
one can study the general case of orbifold projections realizing the orbifold breaking
$\mc{G}\rightarrow\mc{H}\subset\mc{G}$.
This generalization is technically straightforward, at least as far as inner automorphisms are
concerned, and leads to interesting constraints on the allowed fermion configurations.
{}From the usual orbifold field theory point of view,
we have complete freedom of putting at the orbifold fixed points an arbitrary number of $4D$ fermions in
arbitrary representations of the surviving group $\mc{H}$, regardless of the representation
and number of bulk fields. As it will be clear in the following simple example, much of this arbitrariness will be
removed when imposing the orbifold model to admit a resolution.

Consider an $SU(2)$ doublet of Dirac fermions $\psi=(\psi^+,\,\psi^-)^t$ on the $T^2/\Z_2$ orbifold,
satisfying the following orbifold$+$torus boundary conditions:
\be
\begin{array}{c}
\psi(-z)={\mc P}\psi(z)\,,\;\;\;\;\;{\mc P}={\ds e^{\frac{\pi i}2{\sigma_3}}
e^{\frac{\pi i}{2}J_3} }\,,\\
\psi(z+1)=  \psi(z)\,, \\
\psi(z+\tau)= \psi(z)\,,
\end{array}
\label{bc-NA}
\ee
where $J_l$ ($l=1,2,3$) are the generators of $SU(2)$, normalized such that $J_3=\textrm{diag}(1,-1)$.
When applied to the $SU(2)$ gauge fields, the boundary conditions (\ref{bc-NA})
realize the breaking of $SU(2)\rightarrow U(1)$. The components $\psi^\pm$ of $\psi$ satisfy the
conditions given in Eqs.~(\ref{or-cond}) and (\ref{per-cond}), with
$t_{1,2}=0$ and $p=\pm1$, respectively.
It is clear, from the form of the twist matrix ${\cal P}$ in Eq.~(\ref{bc-NA}), that the
resolution of this orbifold configuration requires a background gauge field $A_i$,
at each spherical cap, aligned with the $J_3$ direction: $A_i = A_i^3 J_3$, where
$A_i^3=(\kappa_i/2) \omega_i$, with $\kappa_i=1+4n_i$ and $n_i\geq 0$, so that the bulk
zero mode is present. It is straightforward to solve the Dirac equation for $\psi$
and notice that, aside the usual bulk modes (one left-handed and one right-handed),
$n_i$ left-handed and an equal number of right-handed localized states are present at the fixed point
$z_i$. All left-handed (right-handed) states have +1 (-1) charge under the surviving
$U(1)$ gauge symmetry.
The index of the Dirac operator is zero, as expected, since $\textrm{Tr}[F]=0$ for
$SU(2)$. On the contrary, the most general configuration one can have from the point of view of the
unresolved orbifold theory, would allow to put at the fixed points
an arbitrary number of $4D$ chiral fermions with any charge under the $U(1)$ surviving
gauge group.

It should be said that more fermion configurations can be allowed if one
assumes that extra non-vanishing and non-dynamical
background gauge fields --- under which the fermions are charged --- are
present on the defining orbifold. In this case, the Dirac equation is modified
and one has more allowed fermion configurations, parametrized by the values of the
fluxes associated to the new non-dynamical gauge fields.
As a simple example of this phenomenon, consider again the $SU(2)$ model above.
Assume that another gauge background $A_0$, proportional to the identity, is present, such that
$A_{0,i}=(\kappa_i^{(0)}/2)\omega_i$, with $\kappa_i^{(0)}=4 m_i$
and $m_i\leq n_i$, for simplicity. This background field would not affect the boundary conditions
(\ref{bc-NA}), but it affects the massless Dirac equation. Indeed, including all the
$SU(2)\times U(1)$ gauge background,
one now finds $n_i+m_i$ left-handed and $n_i-m_i$ right-handed localized fermions
at the fixed-point $z_i$, in addition to the same bulk modes as before.
Depending on the values of the integers $m_i$, more fermion configurations are now
allowed.

A less trivial generalization is required to study resolution of fermions in interaction
with other fields. Similarly, an interesting generalization would be to study
the resolutions of the remaining two-dimensional toroidal orbifolds, namely
$T^2/\Z_N$, with $N=3,4,6$.
It would also be interesting to study whether our analysis can be generalized to
supersymmetric theories.

Another important line of development of this work is provided by the study of the localization
of other fields, other than fermions, in the orbifold limit. It would be interesting to study
in particular the gauge field equations of motion and its spectrum of fluctuations around the
background. This will show if, in the orbifold limit, one gets extra scalar states, localized
at the fixed points, other than the expected bulk modes $A_M$. A closely related question would be
to see and possibly identify the scalar fields whose vacuum expectation values fix the size
of the spherical caps we used to resolve the orbifolds. These states are indeed present
in string theory orbifolds, where they always correspond to localized states in the orbifold
limit, namely to so called twisted states.
Equally important would be to see whether, and under what circumstances, the resolved space
solves the gravity equations of motion. We expect that Einstein equations would not
be easily satisfied unless one does not introduce extra fields and/or additional
dynamics.

The analysis performed in this paper might also be useful in better understanding
localized operators arising in orbifold field theories. Among them, particular attention
has been devoted to divergent localized Fayet-Iliopoulos terms in
supersymmetric theories \cite{localizationFI}
and localized tadpoles arising from operators linear in the field strength $F$ of a
non-abelian gauge theory \cite{Tad}. Although to shed some light on how the localized FI
term arises in the orbifold limit would require a study of how to maintain supersymmetry
in our resolved spaces,\footnote{Notice that in many string derived SUSY orbifold models,
the resolution of the orbifold singularities is closely related to the generation of a
localized FI term \cite{dm}.} the latter kind of terms are easier to consider.
Indeed, we have shown that in the resolved space there is a non-trivial classical flux for $F$,
that becomes localized and buried at the fixed points in the orbifold limit, and whose presence
is manifested by the gauge twist matrix $P$. From this point of view, the one-loop induced divergent
tadpoles studied in \cite{Tad} are nothing else that renormalizations of this flux.
Being the latter quantized, it thus implies a renormalization of the radius of the
spherical cap resolving the singularity. In order to show that this is actually the case,
one should couple the system to gravity and see whether a similar divergent tadpole for the
curvature two-form is generated as well.

Similarly, our analysis gives a new twist on the issue of localized anomalies in orbifold
field theories (see \cite{an} for a recent review). The chiral anomaly of a fermion
propagating on a 6D manifold ${\cal M}$ and coupled to a $U(1)$ gauge field is given by
\be
\delta_\alpha \Gamma_6(A) \equiv \alpha \int_{{\cal M}} {\cal A}(x) =
\frac{\alpha}{48\pi^3} \int_{{\cal M}} F^3\,,
\label{ano6D}
\ee
where $\alpha$ is the infinitesimal parameter of the chiral transformation.
In the orbifold limit, the background field strength is given by
\be
\frac{F}{2\pi} = \sum_i \kappa_i \delta^i_2 \,,
\ee
where we have introduced a delta-function two-form, defined as
$\delta^i_2= \delta^2(z-z_i) dz\wedge d\bar z$,
so that the 6D anomaly (\ref{ano6D}) gives rise to a localized 4D anomaly
\be
{\cal A}  = \frac{\alpha}{8\pi^2} \sum_i \kappa_i \, \delta^i_2 \,F^2\,.
\label{ano4D}
\ee
It is by now understood that the localized anomalies of the type (\ref{ano4D}) are
generally regularization dependent, unless there is some extra symmetry that privileges
one particular regularization. A simple instance of such a symmetry might be given by the
permutation of the fixed points, in the case they are indistinguishable from each other.
Requiring this symmetry in the resolved space ${\cal M}$ implies taking the fluxes on the
spherical caps such that $n_1=n_2=n_3=n_4=n$ and $t_1=t_2=0$.
In this case, Eq.(\ref{ano4D}) gives
\be
{\cal A}  = \frac{\alpha}{8\pi^2} \sum_i \bigg(\frac p4 + n\bigg) \, \delta_2^i \,F^2\,,
\label{ano4D-II}
\ee
that corresponds to the sum of the contributions of a bulk fermion
(the factor $p$ in Eq.~(\ref{ano4D-II})) and of the $4n$ localized fermion
fields (the factor $n$ in Eq.~(\ref{ano4D-II})).  If no extra symmetry is
imposed, the localized 4D anomaly (\ref{ano4D-II}) can always be
shifted by an amount that integrates to zero, since the integers
$\kappa_i$ are arbitrary, being only their sum fixed.  There are
however some restrictions. Consider, for instance, the well studied
case of the 5D orbifold $S^1/\Z_2$. If $\tilde\kappa=1$, the only symmetric
choice between the fixed points is $\tilde \kappa_1=\tilde
\kappa_2=1$, resulting in a 5D anomaly equally distributed among the
two fixed-points, as known.
If $\tilde\kappa=0$, the trivial choice $\tilde \kappa_1=\tilde \kappa_2=0$
is not allowed if the integers $\tilde \kappa_i$ must be odd integers, as
we have seen in section 4. The minimal choice in this case is
$\tilde \kappa_1=-\tilde \kappa_2=1$, indicating that a localized, globally
vanishing, anomaly necessarily remains.
Choosing other values of $\tilde \kappa_i$, at fixed $\tilde\kappa=1$ or
$\tilde\kappa=0$, corresponds to
``resolving'' the orbifold in a manner that leads to a different
spectrum of massive fermions on the resolved space. Such states, once
integrated out, gives rise to Chern-Simons couplings whose integrated
anomaly vanishes.

\section*{Acknowledgments}

We would like to thank I. Antoniadis, U. Bruzzo, B. Fantechi, A. Hebecker, M. Quiros,
R. Russo, A. Schwimmer, C.A. Scrucca, L. Silvestrini and F. Zwirner
for interesting discussions and comments.
This research work was partly supported by
the European Community through the RTN
Program ``Across the Energy Frontier'', contract HPRN-CT-2000-00148 and
the RTN network ``The quantum structure of space-time and the
geometric nature of fundamental interactions'', contract HPRN-CT-2000-00131.

\appendix

\section{Chiral fermions on $S^2$}

In this appendix we compute the number and the wave functions in the internal space of the $4D$
chiral fermions which arise from the compactification of a $6D$ chiral fermion $\Psi$
on the two-sphere $S^2$, in presence of a non-vanishing flux for a $U(1)$ field strength $F$.
Although this analysis is standard and well-known (see {\it e.g.} \cite{s2fer}),
it plays a central role in our way of resolving conical singularities by
spherical caps. It is for this reason --- as well as to fix our notations and conventions ---
that is reported here.

The $2D$ Euclidean gamma matrices $\rho^i$ that we use are
\be
\label{2dg}
\rho^1=\sigma_1\,,\;\;\;\;\;\rho^2=\sigma_2\,,\;\;\Rightarrow\;\;\rho_3=i\rho^1\rho^2=-\sigma_3\,,
\ee
where $\sigma_i$ are the standard Pauli matrices.
The gamma matrices $\Gamma^A$ for $6D$ Minkowski space with mostly minus signature can be written as tensor
products of $\rho^i$ with the usual $4D$ ones $\gamma^a$:
\be
\Gamma^a=\gamma^a\otimes\I_2\,,\;\;\;\;\;\Gamma^4=i\gamma_5\otimes\rho^1\,,\;\;\;\;\;
\Gamma^5=i\gamma_5\otimes\rho^2\,,
\ee
where {$\gamma^5=i\gamma^0\ldots\gamma^3$} is the usual $4D$ chirality matrix. The $6D$ chirality matrix
{$\Gamma_7=\Gamma^0\ldots\Gamma^5$} is then the tensor product of the $4D$ and $2D$ chirality matrices:
\be
\Gamma_7=\gamma_5\otimes\rho_3\,.
\ee
A chiral $6D$ fermion $\Psi$ can be decomposed as
\be
\Psi =i \chi_L(x) \otimes \psi_L +\chi_R(x) \otimes \psi_R
\ee
with $\chi$ and $\psi$ four and two dimensional chiral
fermions, respectively.

Since the components along the $4D$ Minkowski space of the $6D$ gauge field and
 vielbein are trivial, the $6D$ massless Dirac equation {$i\Gamma^MD_M\Psi=0$} can be written as
\be
-\dslash_4\chi_L\otimes\psi_L-i\chi_L\otimes\rho^iD_i\psi_L+
i\dslash_4\chi_R\otimes\psi_R+\chi_R\otimes\rho^iD_i\psi_R=0\,.
\ee
In order for $\chi_{L,R}(x)$ to be the components of a $4D$ fermion of mass $m$ the
above equation must be trivially
solved once we impose $i\dslash_4\chi_{L(R)}=m\chi_{R(L)}$. This implies that $\psi_{L,R}$
must satisfy the $2D$ Dirac equation $i\rho^iD_i\psi_{L(R)}=m\psi_{R(L)}$ on the Euclidean extra space.
Massless 4D fermions are in $1$-$1$ correspondence with 2D fermion zero modes in the internal space.
Let us then consider the zero-mode fermion spectrum on the $S^2$.

Since $S^2$ is a non-trivial manifold, it is necessary to introduce
at least two sets to cover it. We introduce spherical coordinates $(\theta_i,\phi_i)$ ($i=1,2$),
with $0 \leq \theta_i < \pi$, $0\leq \phi_i < 2\pi$ on the two sets, so that
$\theta_1=0$ and $\theta_2=0$ represent the North and South poles of $S^2$, respectively.
The two coordinate systems are related by the following transformations:
$\theta_2 = \pi - \theta_1$, $\phi_2 = - \phi_1$.\footnote{Notice that $S^2$ is such a simple
manifold that it is actually not needed to introduce different local coordinates. However,
this way is more suitable when studying the orbifold resolutions where the different spherical
coordinates on the various spherical caps do not have simple relations
between each other.}
The spin connections on the two sets are
\be
\omega_i = (1- \cos\, \theta_i)\, d\phi_i \,,
\label{omega-i}
\ee
and are related by a local $SO(2)\simeq U(1)$ Lorentz transformation: $\omega_2 = \omega_1
- 2 d \phi_1$. By using Stokes theorem, one easily finds that $1/(2\pi) \int_{S^2}R = 2$.
A non-vanishing flux $F$ is obtained by considering a gauge background
proportional to the spin-connection:
\be
A_i = \frac{\kappa}{2} \omega_i \,.
\ee

As usual, the constant $\kappa$ must be an integer, so that electric charges are single valued on $S^2$.
One has $1/(2\pi) \int_{S^2} F = \kappa$ which, by virtue of the Atiyah-Singer index theorem, is
the number of left-moving minus the number of right-moving chiral fermions on $S^2$ in such a $U(1)$
background.
Fermion fields (of unit $U(1)$ charge) in the two sets are related by a
local gauge and Lorentz transformation as follows:
\be
\psi^{(2)}=\ds{e^{i \phi_1 (\kappa + \sigma_3)}}\psi^{(1)}\,.
\label{gauge-Lorentz}
\ee
Accordingly, the covariant derivative on fermions is
\be
D=\partial+iA+\frac{i}2\omega\sigma_3\,.
\ee
The fermion $\psi$ is decomposed as follows in terms of left- and right-moving fields:
\be
\psi^{(i)}=\left(\begin{array}{c}\psi_R^{(i)}\\\psi_L^{(i)}\end{array}\right)\,.
\ee
In both sets, the Dirac equation reads:
\be
\sigma_1 \frac{\partial}{\partial \theta_i} + \frac{\sigma_2}{\sin \theta_i}
\Bigg[ \frac{\partial}{\partial \phi_i}
+\frac i2 (1-\cos \theta_i) (\kappa+\sigma_3) \Bigg] \psi^{(i)} =0 \,.
\ee
An explicit solution is easily found by setting $\psi_{L,R}^{(i)}=\rho_{L,R}^{(i)}(\theta)
e^{i N_{L,R}^{(i)} \phi}$
in both sets, with $N_{L,R}^{(i)}$ integers. Modulo a normalization factor, one finds the
following wave functions:
\be
\left\{
\begin{array}{c}
\psi_L^{(i)} =  (1+\cos \theta_i)^{\frac{N_L^{(i)}+\kappa-1}{2}} \,(1-\cos \theta_i)^{-\frac{N_L^{(i)}}{2}}
e^{iN_L^{(i)}\phi_i}   \\
\psi_R^{(i)} =  (1+\cos \theta_i)^{-\frac{N_R^{(i)}+\kappa+1}{2}}\, (1-\cos \theta_i)^{\frac{N_R^{(i)}}{2}}
e^{iN_R^{(i)} \phi_i}
\end{array}
\right. \,.
\label{spheresolu}
\ee
In order for the solutions to be normalizable on $S^2$ and well defined at $\theta_i=0$, we must require that
\be
N_R^{(i)}\ge0\,,\;\;\;\;\;  N_L^{(i)}\le0 \,,\;\;\;  i=1,2\,.
\label{ang-cond}
\ee
The local gauge+Lorentz transformation (\ref{gauge-Lorentz}) implies
\bea
- N_L^{(2)} = \a\a \kappa-1 + N_L^{(1)} \,, \nn \\
- N_R^{(2)} = \a\a \kappa+1 + N_R^{(1)} \,.
\label{ang-cond2}
\eea
The conditions (\ref{ang-cond}) and (\ref{ang-cond2}) impose severe constraints on the allowed wave functions.
For $\kappa<0$, no left-handed fermions are allowed, whereas one has $|\kappa|$ right-handed fermions with
$0\leq N_R^{(1)} \leq -\kappa -1$; on the contrary, for $\kappa>0$ no right-handed fermions are allowed, whereas
one has $|\kappa|$ left-handed fermions with $1-\kappa \leq N_L^{(1)} \leq 0$.
For $\kappa=0$, no solution is allowed.

\section{Theta functions}

The theta functions with characteristics $a,b=0,1/2$ are defined as follows:
\be
\theta \tw {a}{b}(z|\tau) = \sum_n q^{\frac 12(n+a)^2} e^{2 \pi i (z+b)(n+a)} \,,
\label{theta}
\ee
where $q = \exp (2 \pi i \tau)$. Another common and more compact
notation is the following:
\bea
\theta \tw {1/2}{1/2}=\a\a \theta_1\,; \;\;\;  \theta \tw {1/2}{0}=\theta_2\,; \nn \\
\theta \tw {0}{0}=\a\a \theta_3\,; \;\;\; \theta \tw {0}{1/2}=\theta_4\,.
\eea
These functions are related by the following identities:
\bea
\theta_2(z) = \a\a\theta_1\Big(z-\frac12\Big)\,, \;\;\;\;\;
\theta_3(z) = q^{1/8} e^{-i\pi z} \theta_1\Big(z-\frac12-\frac\tau 2\Big)\,, \nn \\
\theta_4(z) = \a\a-i q^{1/8} e^{-i\pi z} \theta_1\Big(z-\frac\tau2\Big)\,,
\eea
omitting the dependence on the modular parameter $\tau$. They are
clearly holomorphic functions, but
they are not elliptic functions, since they are not exactly periodic.
The function $\theta_1$, for instance, satisfies the following periodicity conditions:
\be
\theta_1 (z+1|\tau)=-\theta_1 (z|\tau)\,, \;\;\;
\theta_1 (z+\tau|\tau)=-q^{-1/2} e^{-2i\pi z}\theta_1 (z|\tau)\,.
\label{theta1-per}
\ee
One of the most important property of theta functions is that they have only one simple
zero and no poles at all inside a fundamental domain ${\cal P}$.
The zeroes $z_i$ of the 4 theta functions $\theta_i$ ($i=1,2,3,4$) are
at $z_1=0$, $z_2=1/2$, $z_3=(1+\tau)/2$ and $z_4=\tau/2$, modulo lattice
shifts.
There are many relations between the $\theta_i$.
The ones we will use in the main text are the following \cite{WW}:
\bea
\!\!\!\!\theta_2^2(z) \theta_4^2(0) = \a\a \theta_4^2(z) \theta_2^2(0)
-\theta_1^2(z) \theta_3^2(0)\,, \;\;\;
\theta_3^2(z) \theta_4^2(0) = \theta_4^2(z) \theta_3^2(0)-\theta_1^2(z) \theta_2^2(0)\,, \nn \\
\!\!\!\!\theta_1^2(z) \theta_4^2(0) = \a\a \theta_3^2(z) \theta_2^2(0)
-\theta_2^2(z) \theta_3^2(0)\,, \;\;\;
\theta_4^2(z) \theta_4^2(0) = \theta_3^2(z) \theta_3^2(0)-\theta_2^2(z) \theta_2^2(0)\,.
\label{theta-relations}
\eea

\end{document}